\documentclass[11pt,preprint]{emulateapj}
\usepackage{natbib}

\usepackage{epstopdf}
\usepackage{color}
\usepackage{epsfig}
\usepackage{comment}
\usepackage{amsmath}
\usepackage{graphicx}
 
\shorttitle{Imaging and Modeling of HERA DATA}

\shortauthors{Carilli, Thyagarajan, Kent, Nikolic, et al.}
 
\begin{document}

\title{Imaging and Modeling Data from the Hydrogen Epoch of Reionization Array}


\author{
C. L. Carilli\altaffilmark{1,2},
N. Thyagarajan\altaffilmark{1},
J. Kent\altaffilmark{2},
B. Nikolic\altaffilmark{2},
K. Gale-Sides\altaffilmark{2},
N.S. Kern\altaffilmark{3},
G. Bernardi\altaffilmark{4,5,6},
A. Mesinger\altaffilmark{7},
S. Matika\altaffilmark{5},
Zara  Abdurashidova\altaffilmark{3},
James E. Aguirre\altaffilmark{8},
Paul Alexander\altaffilmark{2},
Zaki S. Ali\altaffilmark{3},
Yanga  Balfour\altaffilmark{6},
Adam P. Beardsley\altaffilmark{9},
Tashalee S. Billings\altaffilmark{8},
Judd D. Bowman\altaffilmark{9},
Richard F. Bradley\altaffilmark{1},
Phil  Bull\altaffilmark{10},
Jacob  Burba\altaffilmark{11},
Carina  Cheng\altaffilmark{3},
David R. DeBoer\altaffilmark{3},
Matt  Dexter\altaffilmark{3},
Eloy  de~Lera~Acedo\altaffilmark{2},
Joshua S. Dillon\altaffilmark{3},
Aaron  Ewall-Wice\altaffilmark{12},
Nicolas  Fagnoni\altaffilmark{1},
Randall  Fritz\altaffilmark{6},
Steve R. Furlanetto\altaffilmark{13},
Kingsley  Gale-Sides\altaffilmark{2},
Brian  Glendenning\altaffilmark{1},
Deepthi  Gorthi\altaffilmark{3},
Bradley  Greig\altaffilmark{14},
Jasper  Grobbelaar\altaffilmark{6},
Ziyaad  Halday\altaffilmark{6},
Bryna J. Hazelton\altaffilmark{15,16},
Jacqueline N. Hewitt\altaffilmark{12},
Jack  Hickish\altaffilmark{3},
Daniel C. Jacobs\altaffilmark{9},
Alec  Josaitis\altaffilmark{2},
Austin  Julius\altaffilmark{6},
Joshua  Kerrigan\altaffilmark{11},
Honggeun  Kim\altaffilmark{12},
Piyanat  Kittiwisit\altaffilmark{9},
Saul A. Kohn\altaffilmark{8},
Matthew  Kolopanis\altaffilmark{9},
Adam  Lanman\altaffilmark{11},
Paul  La~Plante\altaffilmark{8},
Telalo  Lekalake\altaffilmark{6},
Adrian  Liu\altaffilmark{17},
David  MacMahon\altaffilmark{3},
Lourence  Malan\altaffilmark{6},
Cresshim  Malgas\altaffilmark{6},
Matthys  Maree\altaffilmark{6},
Zachary E. Martinot\altaffilmark{8},
Eunice  Matsetela\altaffilmark{6},
Mathakane  Molewa\altaffilmark{6},
Miguel F. Morales\altaffilmark{15},
Tshegofalang  Mosiane\altaffilmark{6},
Abraham R. Neben\altaffilmark{12},
Juan Mena Parra\altaffilmark{12},
Aaron R. Parsons\altaffilmark{3},
Nipanjana  Patra\altaffilmark{3},
Samantha  Pieterse\altaffilmark{6},
Jonathan C. Pober\altaffilmark{11},
Nima  Razavi-Ghods\altaffilmark{2},
James  Robnett\altaffilmark{1},
Kathryn  Rosie\altaffilmark{6},
Peter  Sims\altaffilmark{11},
Angelo  Syce\altaffilmark{6},
Peter K.~G. Williams\altaffilmark{18},
Haoxuan  Zheng\altaffilmark{12}
}

\altaffiltext{1}{National Radio Astronomy Observatory, P. O. Box 0,Socorro, NM 87801, USA; ccarilli@nrao.edu, ORCID: 0000-0001-6647-3861}                                                     
\altaffiltext{2}{Astrophysics Group, Cavendish Laboratory, JJ Thomson Avenue, Cam
bridge CB3 0HE, UK}

\altaffiltext{3}{Department of Astronomy, University of California, Berkeley, CA}

\altaffiltext{4}{INAF-Istituto di Radioastronomia, via Gobetti 101, 40129, Bologna, Italy}

\altaffiltext{5}{Department of Physics and Electronics, Rhodes University, PO Box 94, Grahamstown, 6140, South Africa}

\altaffiltext{6}{The South African Radio Astronomy Observatory (SARAO), 2 Fir Street, Black River Park, Observatory (North Gate entrance), 7925, South Africa}

\altaffiltext{7}{Scuola Normale Superiore, 56126 Pisa, PI, Italy}

\altaffiltext{8}{Department of Physics and Astronomy, University of Pennsylvania, Philadelphia, PA}

\altaffiltext{9}{School of Earth and Space Exploration, Arizona State University, Tempe, AZ}

\altaffiltext{10}{Queen Mary University of London, London, UK}

\altaffiltext{11}{Department of Physics, Brown University, Providence, RI}

\altaffiltext{12}{Department of Physics, Massachusetts Institute of Technology, Cambridge, MA}

\altaffiltext{13}{Department of Physics and Astronomy, University of California, Los Angeles, CA}

\altaffiltext{14}{School of Physics, University of Melbourne, Parkville, VIC 3010, Australia}

\altaffiltext{15}{Department of Physics, University of Washington, Seattle, WA}

\altaffiltext{16}{eScience Institute, University of Washington, Seattle, WA}

\altaffiltext{17}{Department of Physics and McGill Space Institute, McGill University, 3600 University Street, Montreal, QC H3A 2T8, Canada}

\altaffiltext{18}{Harvard-Smithsonian Center for Astrophysics, Cambridge, MA}

\begin{abstract}

We analyze data from the Hydrogen Epoch of Reionization Array. This is the third in a series of papers on the closure phase delay-spectrum technique designed to detect the HI 21cm emission from cosmic reionization. We present the details of the data and models employed in the power spectral analysis, and discuss limitations to the process. We compare images and visibility spectra made with HERA data, to parallel quantities generated from sky models based on the GLEAM survey, incorporating the HERA telescope model. We find reasonable agreement between images made from HERA data, with those generated from the models, down to the confusion level. For the visibility spectra, there is broad agreement between model and data across the full band of $\sim 80$MHz. However, models with only GLEAM sources do not reproduce a roughly sinusoidal spectral structure at the tens of percent level seen in the observed visibility spectra on scales $\sim 10$ MHz on 29 m baselines. We find that this structure is likely due to diffuse Galactic emission, predominantly the Galactic plane, filling the far sidelobes of the antenna primary beam. We show that our current knowledge of the frequency dependence of the diffuse sky radio emission, and the primary beam at large zenith angles, is inadequate to provide an accurate reproduction of the diffuse structure in the models. We discuss implications due to this missing structure in the models, including calibration, and in the search for the HI 21cm signal, as well as possible mitigation techniques.

\end{abstract}


\section{Introduction}

Cosmic reionization corresponds to the epoch when the UV radiation from the first luminous sources (stars, black holes), reionizes the neutral Hydrogen that pervaded the Universe after cosmic recombination. Measurements of quasar absorption lines, Ly$\alpha$ galaxy demographics, and the cosmic microwave background, have constrained the redshift at which the ionization fraction reaches 50\% in the intergalactic medium (IGM), to be $z = 8.1\pm 1$, with a duration (from 25\% to 75\% ionized), of $\Delta z \sim \pm 1$ \citep{greig17}. While the basic epoch and duration of reionization are reasonably well constrained, many important questions remain about the process of reionization (eg. inside-out or outside-in?), and the sources of reionization (eg. small galaxies? big galaxies? low to intermediate mass black holes?).

It is widely recognized that the 21cm line of neutral hydrogen is a potentially powerful probe of the physics of cosmic reionization \citep{furl13, fan06, morales10}. Imaging at low radio frequencies (100 MHz to 200 MHz), has the potential to determine the large scale structure of the Universe, as dictated by the combined processes of dark matter evolution and reionization. Numerous interferometric experiments are currently operating with the goal of making the first statistical (ie. power-spectral), detection of the HI 21cm signal from large scale structure in the Universe during cosmic reionization \citep{deboer17, patil17, beards16, barry19, li19, kolopanis19, trott20}.

A major hurdle to making the first HI 21cm detection remains removal of the strong foreground radio continuum emission, corresponding to radio synchrotron emission from the Milky Way, and from distant radio galaxies. The foregrounds have a mean surface brightness four to five orders of magnitude larger than the HI 21cm signal, even in the quietest parts of the sky.  Different experiments are employing varied techniques to obtain this first detection. The orignally proposed technique, outlined in eg. \citet{tegmark97, morales05, harker10, harker09} (see the reviews in \citet{furlanetto06, morales10, zaroubi12}), employs calculating the three dimensional power spectrum from image cubes, where the three space dimensions in the image cube (RA, Dec, and frequency, the latter of which corresponds to distance via the redshift of the HI 21cm line), transform to the conjugate wavenumber ($k$) in power spectrum space. The image cubes are generated from the interferometric data via the standard Fourier transform relationship between visibilities and sky-plane surface brightness. The radio continuum emission is subtracted via multiple processes, including identification of point sources in the image domain, then `peeling' these sources in the uv-plane \citep{noordam04}, as well as subtracting smooth-spectrum models fit to the image cubes, or visibilities, themselves \citep{chapman13, zald04}. Variants of these techniques have been employed in the recent analysis of LOFAR data by \citep{patil17}, who calculate both the cylindrical (ie. line-of-sight; see below), and the spherical (ie. three-dimensional), power spectrum. At the other extreme is the `delay spectrum' approach, employing the relationship between frequency and cosmic distance (ie. redshift), to obtain a power spectrum of the HI 21cm signal from the Fourier transform of interferometric visibility spectra along the frequency axis. In this case, since the baseline is fixed, there will be spatial `mode-mixing' as a function of frequency, but this effect is minor on short spacings and moderate frequency ranges. In this `delay space' (where delay is the Fourier conjugate of frequency), the smooth spectrum foreground continuum emission is naturally limited to small delays ($\sim k_\parallel$-modes), although the real and imaginary parts of a given visibility will have frequency structure due to continuum sources not at the phase tracking center \citep{datta10, parsons14, morales19}.

In a series of papers, we are presenting a new approach to the HI 21cm detection, namely, using the closure phase spectra to obtain a power-spectral detection of the HI 21cm emission from reionization. Our technique parallels the delay spectrum approach discussed above, where the smooth spectrum continuum emission is limited to the small delay (or $k_\parallel$) modes, but, as opposed to using interferometric visibility spectra, we employ closure phase spectra.  The closure phase approach has the distinct advantage of being independent of antenna-based complex gains, and hence is robust to calibration terms that are separable into antenna-based contributions \citep{TMS18}.  For the basics of closure phase and our power spectral technique, we refer the reader to: \citet{thya18, carilli18a}.In very brief, the closure phase corresponds to the phase of the triple-product, or `bi-spectrum', of the three complex visibilites measured from three antennas that form a triangle in the array. The closure phase has the important property that the phases introduced by the electronics, and the atmosphere, to each element of the array, cancels in the triple product, such that the closure phase represents a true measure of the attributes of the sky signal, independent of standard antenna-based calibration terms. This interesting property was recognized early in the history of interferometry \citep{jennison58}, and has long been used in radio and optical interferometry to infer properties of the sky brightness, in situations where determining antenna-based gains is difficult. For the mathematics, see \citet{thya18, carilli18a}.

This is the third in our paper series, in which we present the detailed data and modeling that is then employed in the application of the closure phase technique to HERA data presented in \citet{nithya19a}, while the initial mathematical foundations for comprehending the bispectrum phase in the context of EoR power spectrum has been detailed in \citet{nithya19b}. We employ data from the first season of observations by the Hydrogen Epoch of Reionization Array (HERA), for a 50 element array.  We focus on two fields. First is the field around the transit of the strong radio galaxy, Fornax A.  This field provides a number of distinct advantages in testing the closure phase spectral approach to HI 21cm power spectral estimation. The second is a cold region of the sky with no dominant sources, at J0137-3042. 

We present and characterize the data employed, and compare the measured interferometric visibilities, and resulting images, with a detailed modeling of the sky and telescope.  Modeling of the sky and telescope is a crucial component of the power spectral analysis, providing the basis of comparison of the measured power spectra to those expected from the foregrounds, the noise, and the HI 21cm signal \citep{nithya19a}. The imaging results are a text-book example of strongly confusion limited imaging in radio interferometry.

For the visibility spectra, we find reasonable broad agreement between data and models over the 80 MHz band, but the models using only point source models from the GLEAM survey  \citep{HW17} misses signicant spectral structure on scales $\sim 10$ MHz. We show that this excess spectral structure is likely due to diffuse Galactic emission missing from the GLEAM models. We summarize the potential implications of this missing structure on HERA data analysis, and possible mitigation techniques. 

\section{Hydrogen Epoch of Reionization Array}
\label{sec:HERA}

HERA is an interferometric array designed with the purpose of optimizing the search for the HI 21cm fluctuations during cosmic reionization using a delay-spectrum approach \citep{deboer17}. The array is currently under construction, with a goal of having a 331 element array of 14m diameter parabolic antennas, distributed in a hexagonal grid pattern, with grid spacing separated by 14.6m. The antennas are not steerable -- the array is a `zenith-only' instrument, at a latitude of $-30.7^\circ$. The primary beam FWHM at 150 MHz is $8.3^\circ$, with a maximum baseline for the 331 element array of about 300m. Another twenty elements will be deployed out to maximum baselines of 1km. 

In this paper we analyze data from the months of February to March, 2018. We employ the 18 days that comprise HERA Internal Data Release 2.1 \citep[IDR2,][]{dillon18}. The array at this time consisted of a partial hexagonal array of 50 antennas (HERA-50), with 10.7 second averaging time. The layout of the array used in this analysis can be seen in Figure 2 in \citet{carilli18a}.  These data have been inspected for quality assurance.  The spectral data have 1024 channels from 100 MHz to 200 MHz, with a channel width of 97 kHz, and full linear polarization.

The data for imaging have been flagged using the standard HERA procedures \citep{kerrigan19}. The data have been calibrated using a hybrid process of initial sky-based delay calibration, then redundant baseline calibration, then a sky-based calibration procedure to determine the missing parameters inherent in the redundant baseline calibration process, and to determine the absolute gain scale and bandpass \citep{kern19, dillon17}.

In the imaging analysis below, we employ the calibrated IDR2 data for imaging, and amplitude and phase spectral plots. The data have been LST-binned over 18 days, meaning each record at a given LST has been averaged over the 18 days to create a single uv-dataset.  We analyze data around the transit of the strong extragalactic radio source, Fornax A (RA = 03:22), and around a cold region of the sky at J0137-3042. For the best image presented below of the Fornax field, we also employ a bandpass self-calibration process using a CLEAN component model generated from the data. Self-calibration in the case of Fornax A was required due to the dynamic range issues posed by the large flux density of Fornax A. The standard bandpass calibration process for HERA is weighted toward sources near the pointing center (= zenith), in the calibration fields \citep{kern19}). Fornax A, being well down in the primary beam, has a singificant residual spectral shape imposed by the primary beam shape as a function of frequency. Hence, prior to self-calibration, the residual sidelobes from Fornax A are large, and essentially swamp most of the fainter emission in the field. After bandpass self-calibration using Fornax A itself, these sidelobes are greatly mitigated. Of course, the spectral shape for other sources in the field is not conserved, but that is less relevant for these typically 10 to 100 times fainter sources, in the final broad band continuum image. However, when analyzing visibility spectra in the amplitude and phase, we plot the original calibrated IDR2 data without bandpass self-calibration. Self-calibration was not employed in the J0137-3042 field.

Imaging is performed with CASA CLEAN, using a multi-frequency synthesis \citep[MFS;][]{rau11}, from 120 MHz to 180 MHz, and Briggs weighting of the visibilities with a robust parameter of -0.5. This weighting results in a synthesized beam of FWHM = $45' \times 35'$, major axis position angle = $65^\circ$, at the effective frequency of 150 MHz. We have explored MFS using between 1 and 3 Taylor terms in the imaging, and find a small improvement using the higher order. The peak sidelobe of the synthesized beam using the broad band MFS is $\sim 20$\%.

In the analysis below, all flux densities, noise values, and related, are based on the measured brightnesses in the resulting images.

\section{Models vs. Data}
\label{sec:model_vs_data}

We selected two fields to explore the visibilities and imaging of the data in comparison to the sky and telescope modeling, in the context of presenting the data that is then used in our closure phase power spectral analysis \citep{nithya19a}. One field contains a strong, relatively compact source, Fornax A. This field has some interesting characteristics in terms of diagnostics of the closure phase spectra \citep{carilli18a}. The second is a typical field in a quiet region of the sky.

\subsection{The Fornax~A Field}

Figure~\ref{fig:f1} is a full sky radio image at 408 MHz \citep{haslam}.  Fornax A is at J0322-3712. Fornax A is situated in one of the coldest regions of the low frequency sky, with a mean brightness temperature $\sim 180$ K at 150 MHz \citep{costa08}. 

The blue and green lines indicate the horizon for the HERA array for the Fornax field and for the J0137 field discussed below. Notice that large portions of the Galactic plane are always above the horizon for HERA, even for transit observations of the coldest regions of the sky. We return to this point below.

Fornax A is a bright radio source, comprised of two steep spectrum radio lobes, with a full angular extent of the outer boundaries of the radio lobes $\sim 55'$. Hence, Fornax A is marginally resolved in the HERA data presented herein  (resolution of $\sim 40'$). Fornax A has a total flux density at 154 MHz of $750 \pm 142$ Jy, and an integrated low frequency spectral index $\sim -0.8$ \citep{mckinley15}.

Figure~\ref{fig:f2} shows the HERA image of the Fornax~A field at time of transit, after bandpass self-calibration. We fit a Gaussian model to Fornax~A in the HERA image and obtain a total flux density of 173~Jy at the mean frequency of 150 MHz, a peak surface brightness of 120 Jy beam$^{-1}$, and a nominal deconvolved source FWHM of $36'\times 18'$. The implication is that Fornax~A at transit is at the 23\% power-point of the HERA primary beam.  This value is roughly consistent with the primary beam response at the position of Fornax~A ($6.5^\circ$ from the zenith at transit; \citet{fag17,nun19}).  Being well down in the HERA primary beam, Fornax~A makes only a minor contribution to the system temperature ($\sim 12$~K at 150~MHz at transit, see below). We note that the next brightest source in the HERA beam is $\sim 8$~Jy. 

Having a dominant and relatively compact source in the field has a number of distinct advantages when exploring the closure phase spectral approach to detecting the HI 21cm power spectrum from cosmic reionization. The dominant compact source drives the closure phase to zero, and only small fluctuations, much less than a radian, remain.  In \citet{carilli18a, kent18}, we have shown that the closure phase spectra converge on zero across the $\sim \pm 20$ minutes of the transit of Fornax A. Further, in \citet{kent18}, we show the redundancy of the closure spectra across redundant triads becomes substantially better when Fornax dominates the visibilities. Of course, having a dominant compact source is not a  fundamental requirement in the closure phase delay spectrum search for the HI 21cm signal, as was shown in \citet{thya18}, in which more general foreground models were assumed. In section 3.2 we explore a more general quiet-sky field in both imaging and spectral domain. 

We build a model for the Fornax~A field, from the GLEAM low frequency survey \citep{HW17}. The GLEAM model includes source flux densities at 150MHz, plus a spectral index determined by the GLEAM survey. We add about 8000 point sources from the GLEAM survey over a $30^\circ$ diameter area. These correspond to all the sources in the GLEAM catalogue over the full region, down to the GLEAM flux density limit of 50 mJy at 154~MHz (5$\sigma$). The one exception is Fornax A itself, which is not in the GLEAM catalog, and is clearly a very spatially extended radio source. For Fornax A, we used a separate model \citep[private communication with Patricia Carroll and Ruby Byrne;][]{carroll16}, based on MWA observations from 2013, now in the public archive.

We employ the Precision Radio Interferometry Simulator \citep[PRISim\footnote{PRISim is publicly available for use under the MIT license at \href{https://github.com/nithyanandan/PRISim}{https://github.com/nithyanandan/PRISim}};][]{PRISim_software} to generate visibilities from the model similar to that in \citet{thya15}.  We adopt a model for the array using the same antennas that were used during the observations (HERA-50). For the primary beam, PRISim uses the power pattern determined from electromagnetic modeling of the HERA 14~m antenna \citep{fag17}. We generate a non-tracking visibility data set for $\pm 10$ min around the transit of Fornax A.  

We generate visibilities with and without thermal noise. For the purpose of the imaging and visibility model comparisons presented herein, we employ the noiseless data. We show below that the images are confusion limited relative to the expected thermal noise level by more than two orders of magnitude. Including thermal noise makes no discernible difference to the model image results. The thermal noise becomes relevant in the power spectral analysis, as the ultimate limit to detection of the HI 21cm signal, and we consider thermal noise for HERA in detail in \citet{nithya19a}.

PRISim generates a transit dataset in HD5 format. These data are then converted to FITS format, and fringe tracked at the zenith and RA of the first record. In the imaging, we employ $\pm 2$~min around transit of Fornax~A. The FITS uv-data is converted to a CASA measurement set using pyuvdata tools \citep{pyuvdata}. The same imaging parameters are then employed for the model data as for the observed data.

Figure~\ref{fig:f2} shows the resulting image of the Fornax~A field from the data (color scale), and the GLEAM model (contours). In this case, we did not include Fornax~A itself in the simulated model, to better show the results for the fainter sources in the field. Figure~\ref{fig:f2} also shows the difference image, derived by subtracting the model and data images. 

Figure~\ref{fig:f2} shows good agreement between the model image and the observed image.  The measured rms of the surface brightness fluctations outside the primary beam in this image is 0.4~Jy~beam$^{-1}$, while within the primary beam, the rms of the surface brightness fluctuations is about twice this value. Note that these images have not been corrected for the primary beam power response. Hence, the rms noise measured within the primary beam represents confusion noise due to faint sources that fill each synthesized beam (see below). Outside the primary beam, the sky sources are highly attenuated, and the measured noise represents sidelobe confusion noise. For the brighter sources, the flux densities at matched resolution typically agree to better than 10\%. This agreement is comparable to the GLEAM absolute flux density scale uncertainty of 8\%, in the relevant declination range \citep{HW17}.
 
The theoretical thermal noise for this HERA-50 image is less than 1~mJy \citep{carilli18b, parsons17}. The measured rms on the image is four hundred times larger. The low resolution of HERA-50 implies that the resulting images are strongly in-beam source confusion limited.  Quantitatively, the synthesized beam area is about 0.5 deg$^2$.  The average areal density of GLEAM sources down to 50~mJy is 12 sources per deg$^2$ at 154~MHz. This implies, on average, six sources brighter than 50~mJy within every synthesized beam of HERA-50, or typically at least 0.3~Jy total flux density per synthesized beam. In other words, thermal noise is not discernible on HERA images. Every synthesized beam is dominated by sources at a level orders of magnitude larger than the noise. 

Figure~\ref{fig:f3} shows visibility spectra for one record at Fornax transit, from the IDR2 calibrated data, compared to simulated PRISim visibilities, on the three 29~m baselines that make up an equilateral closure triad in the array. Figure~\ref{fig:f4} shows the corresponding closure phase spectrum for this triad. The general shape and magnitudes are similar, at the $\sim 10\%$ level, with the exception that the HERA data spectra show more structure on scales $\sim 10$ MHz, than the model. We investigate this extra structure below.

\subsection{The J0137-3042 Field}

The J0137-3042 field is typical of a high Galactic latitude field with no dominant sources.  We have modeled this field using the GLEAM catalog, as per the Fornax field modeling above, with about 8000 GLEAM sources included over the $30^\circ \times 30^\circ$ area. The model was then employed, along with a HERA telescope model, to generate a visibility data set using both the PRISim software. Images were generated using CASA with parameters as given in Section~\ref{sec:HERA}.  

Figure~\ref{fig:f5} shows the resulting images for the data itself (color-scale) and the model (contours). In this case, no self-calibration was required to reach the confusion noise level. As with the Fornax field, the agreement is very good, down to the confusion level of $\sim 0.4$~Jy~beam$^{-1}$. Figure~\ref{fig:f5} also shows a difference image between the model and the data. Residuals are predominantly at the noise level, besides at the position of a few of the bright sources, where differences are again $\sim 10\%$.  

The residual image does show some large-scale structure at the $\sim 1\sigma$ surface brightness level, appearing as rough, broad stripes of positive and negative contours, extending from the northeast to the southwest, across the primary beam. This large scale residual may indicate diffuse Galactic emission in the real data that is not included in the GLEAM model. To test this idea, we include in Figure~\ref{fig:f5}, the reprocessed Haslam 408 MHz total power image image of the J0137-3042 field \citep{haslam, remazeilles15}. Evident on the image are large structures oriented along a similar direction, and of similar scale, as the large scale residuals seen in the HERA difference image. Note that an interferometer has no sensitivity to the total power in the image, and indeed, to any structures larger than the fringe of the shortest spacing of the array. In our case, this corresponds to scales larger than about 5$^\circ$. Given such diffuse emission, missing short spacings in an interferometric image will lead to positive and negative artifacts on scales comparable to, or larger than, the shortest fringe spacing of the array, as appears to be the case in the difference image.

Figure~\ref{fig:f6} shows visibility spectra (amplitude and phase) for three baselines that make up an equilateral 29~m triad, for both the data and the model. Figure~\ref{fig:f7} shows a similar comparison of the closure spectra for the 29~m equilateral triad.  There is good agreement between the large scale spectral structures between model and data. However, again, the visibility amplitude and phase spectra show considerable smaller-scale structure, in particular a roughly sinusoidal pattern on frequency scales of $\sim 10$~MHz across the band, with amplitudes of $\pm 20\%$ to 50\%. This structure is not seen in the model spectra.

\section{Modeling limitations}

\begin{figure}
\plotone{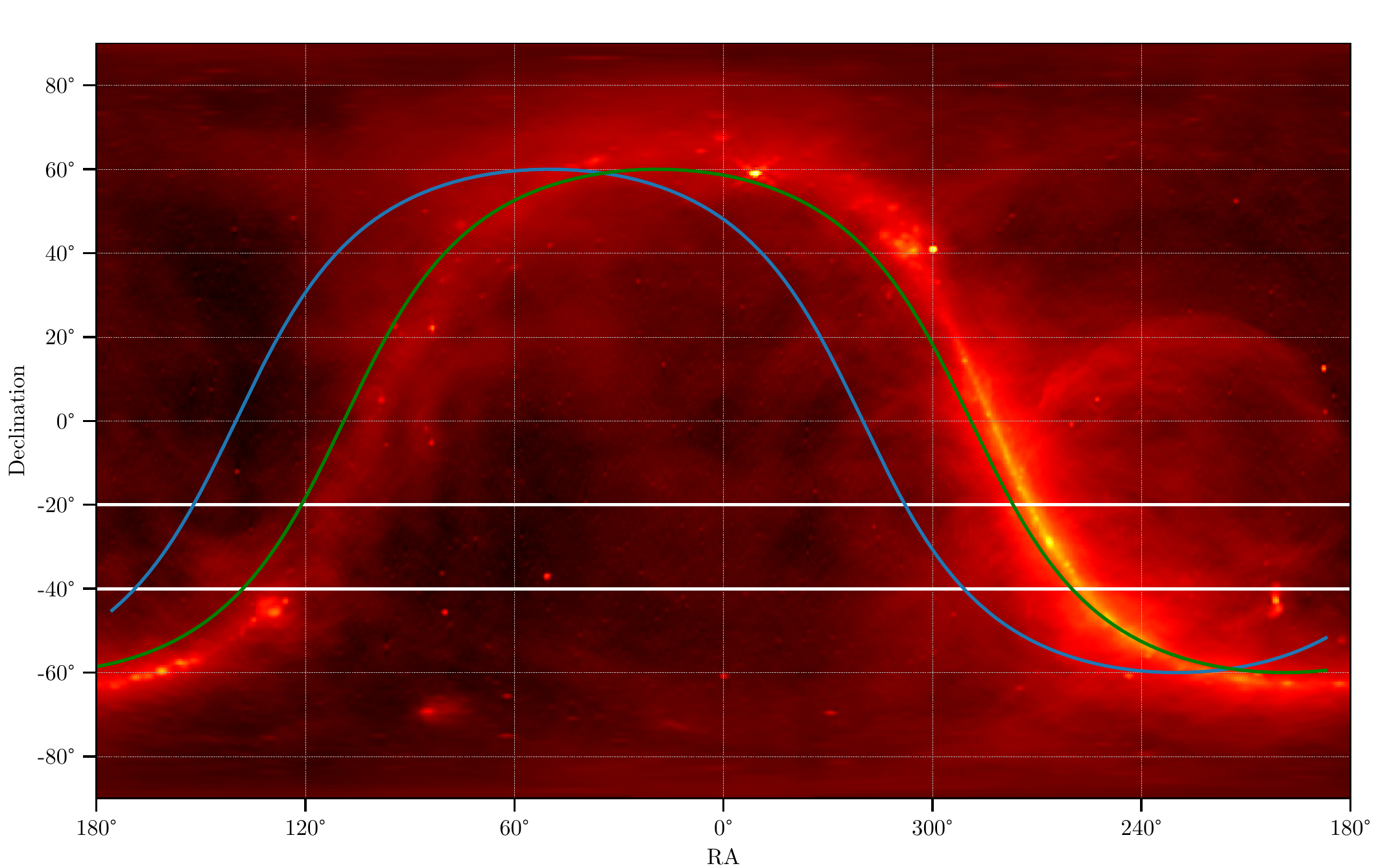}
\caption{All sky radio image at 408 MHz from the Haslam survey \citep{haslam}. The solid white lines indicate $\pm 10^\circ$ north and south of zenith ($\sim $ FWHM of the primary beam at 125~MHz). The blue line shows the horizon at the HERA site, for an LST centered on the Fornax field at transit (Fornax A can be seen at RA = +53$^\circ$, Dec = $-27^\circ$). The green line shows the horizon at transit for the J0137-3042 field. 
}
\label{fig:f1}
\end{figure}

\begin{figure}
\includegraphics[scale=0.65]{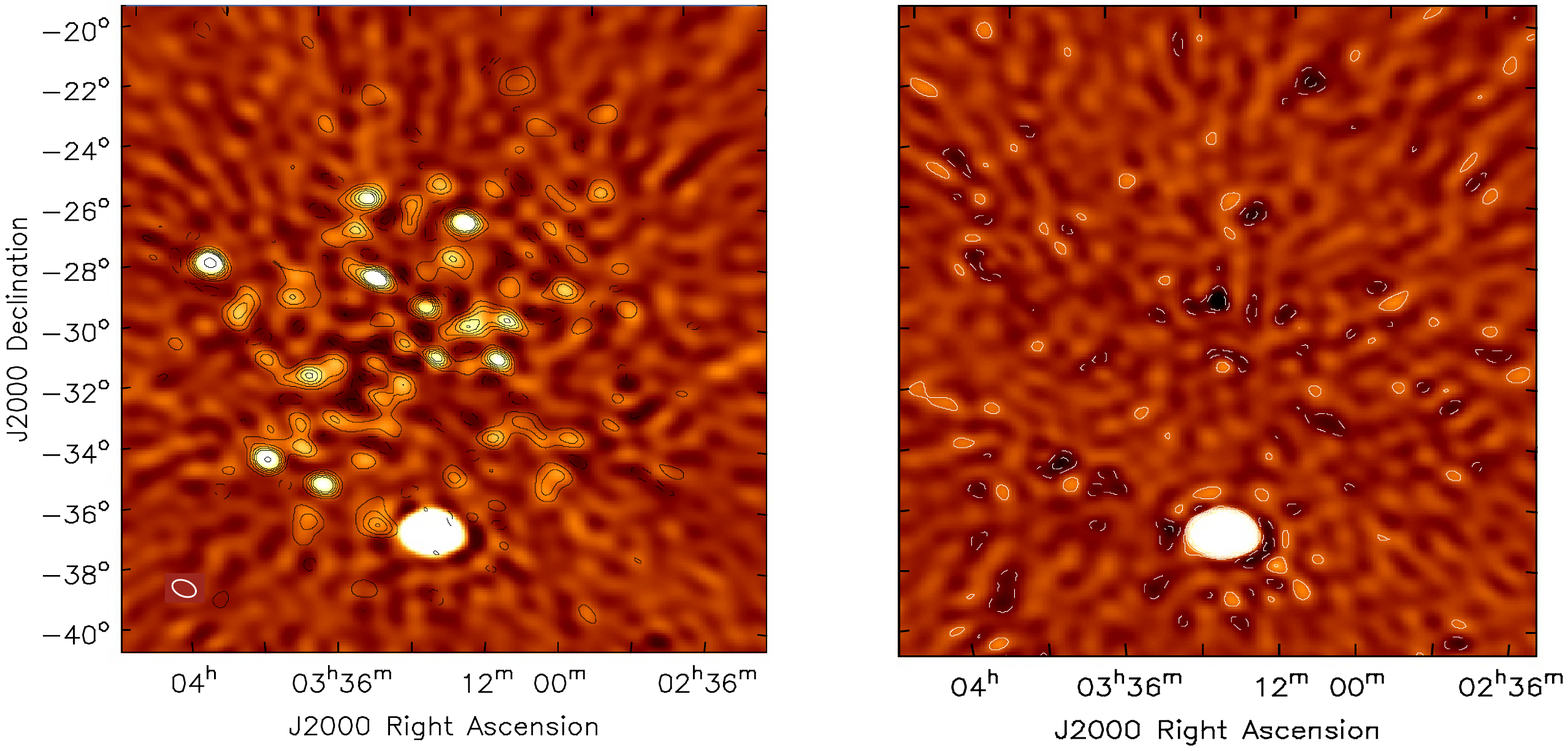}
\caption{Left: Image of the Fornax field using 4~min of data over transit. The color scale shows the image made from the LST-binned IDR2 data from HERA, involving 18 days, using a multifrequency synthesis from 120 MHz to 180 MHz. The contours show the Fornax field GLEAM model, passed through the PRISim simulation and HERA telescope model, to generate visibilities, then imaged in CASA in the exact same way as for the real data. The GLEAM model in this case does not include Fornax A itself, to better show the underlying distribution of fainter sources. The contour levels are: -1.2, -0.6, 0.6, 1.2, 1.8, 2.4, 3.0 Jy beam$^{-1}$, and the resolution is $43' \times 33'$, PA = $65^\circ$. Dashed contours are negative. The rms noise level outside the main beam is 0.4~Jy~beam$^{-1}$. Right: Difference image between model and data, with the same contour levels, to indicate quantitatively the relative magnitude of the residual features. 
}
\label{fig:f2}
\end{figure}

\begin{figure}
\plottwo{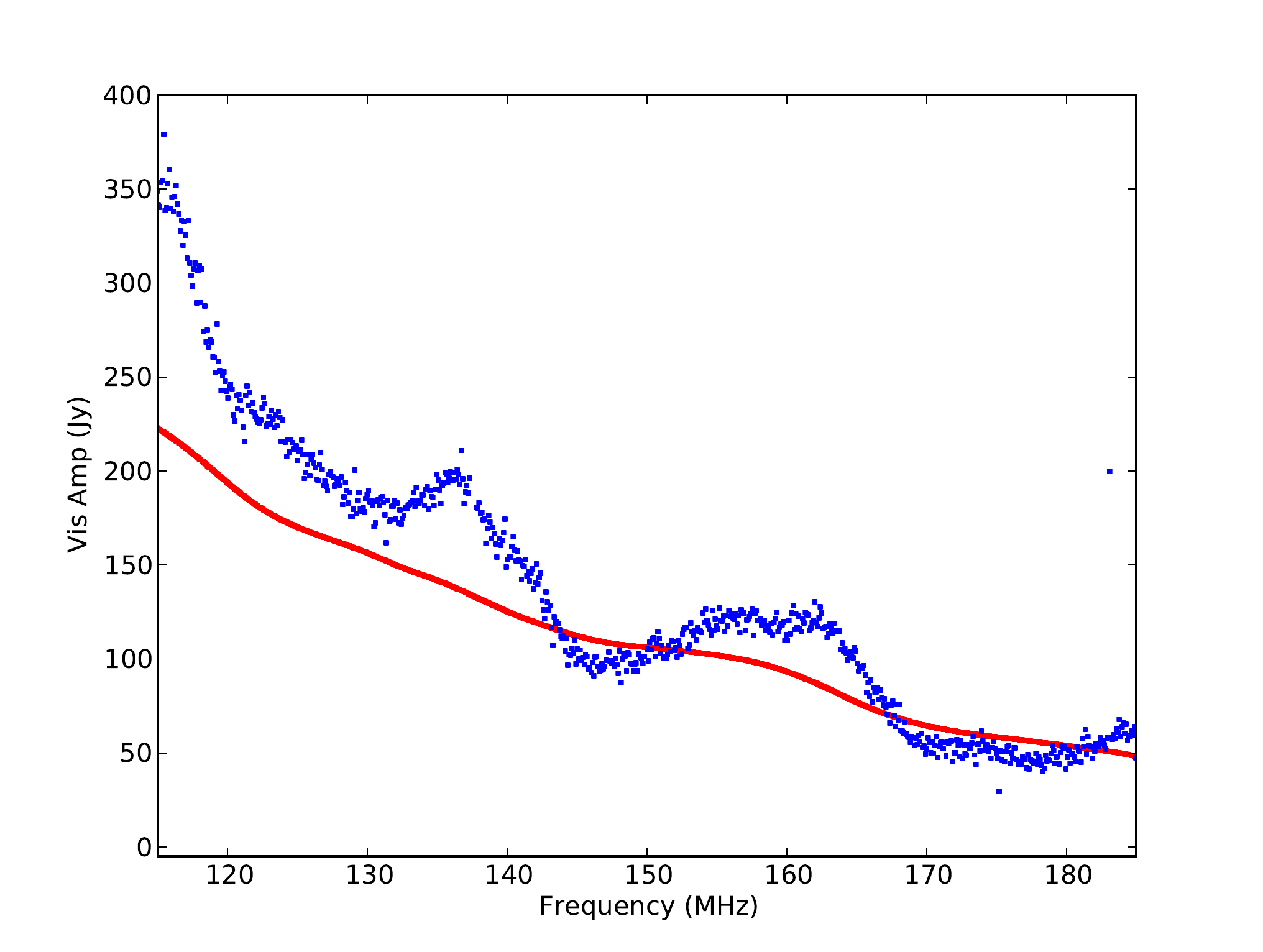}{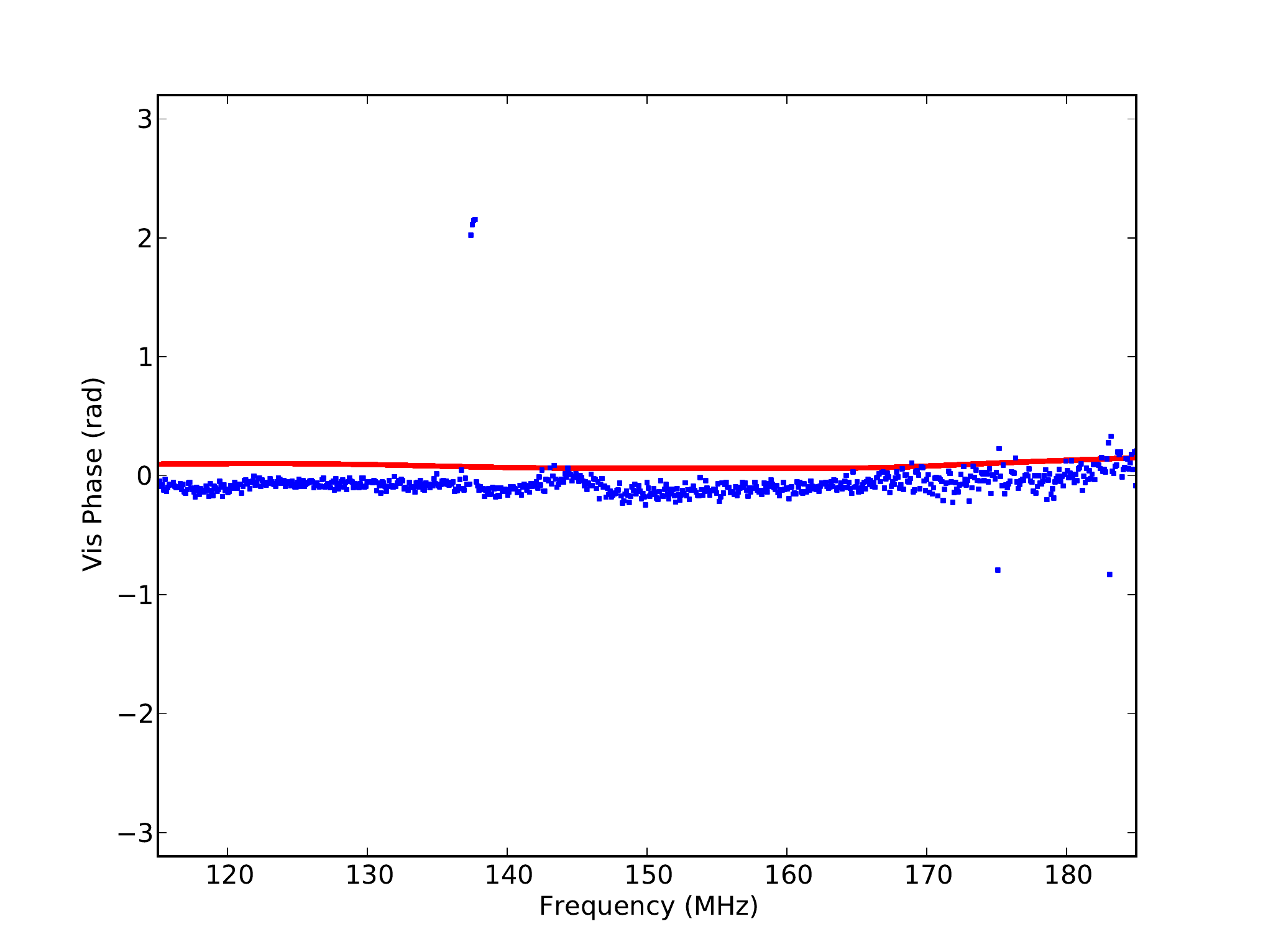}
\plottwo{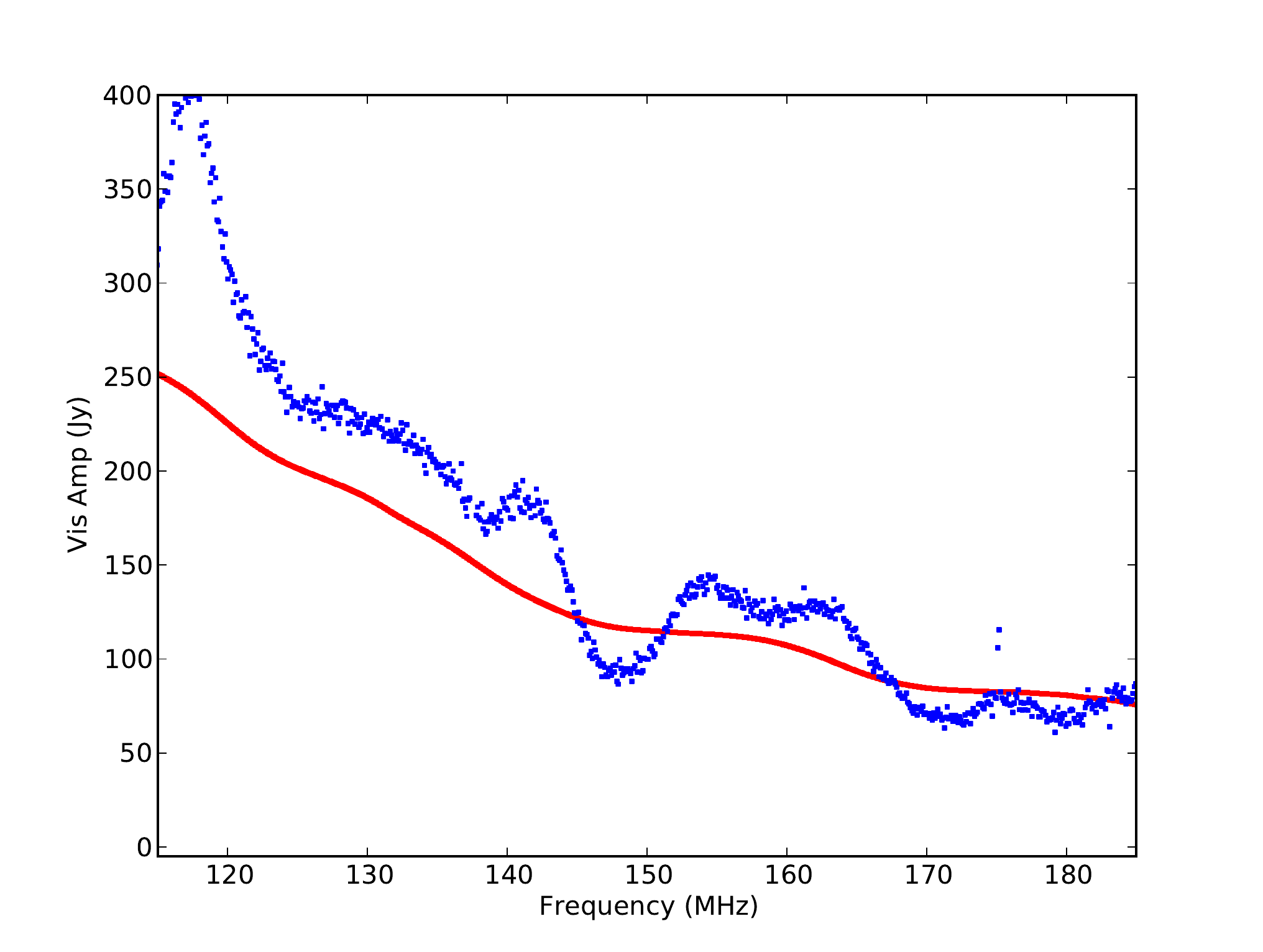}{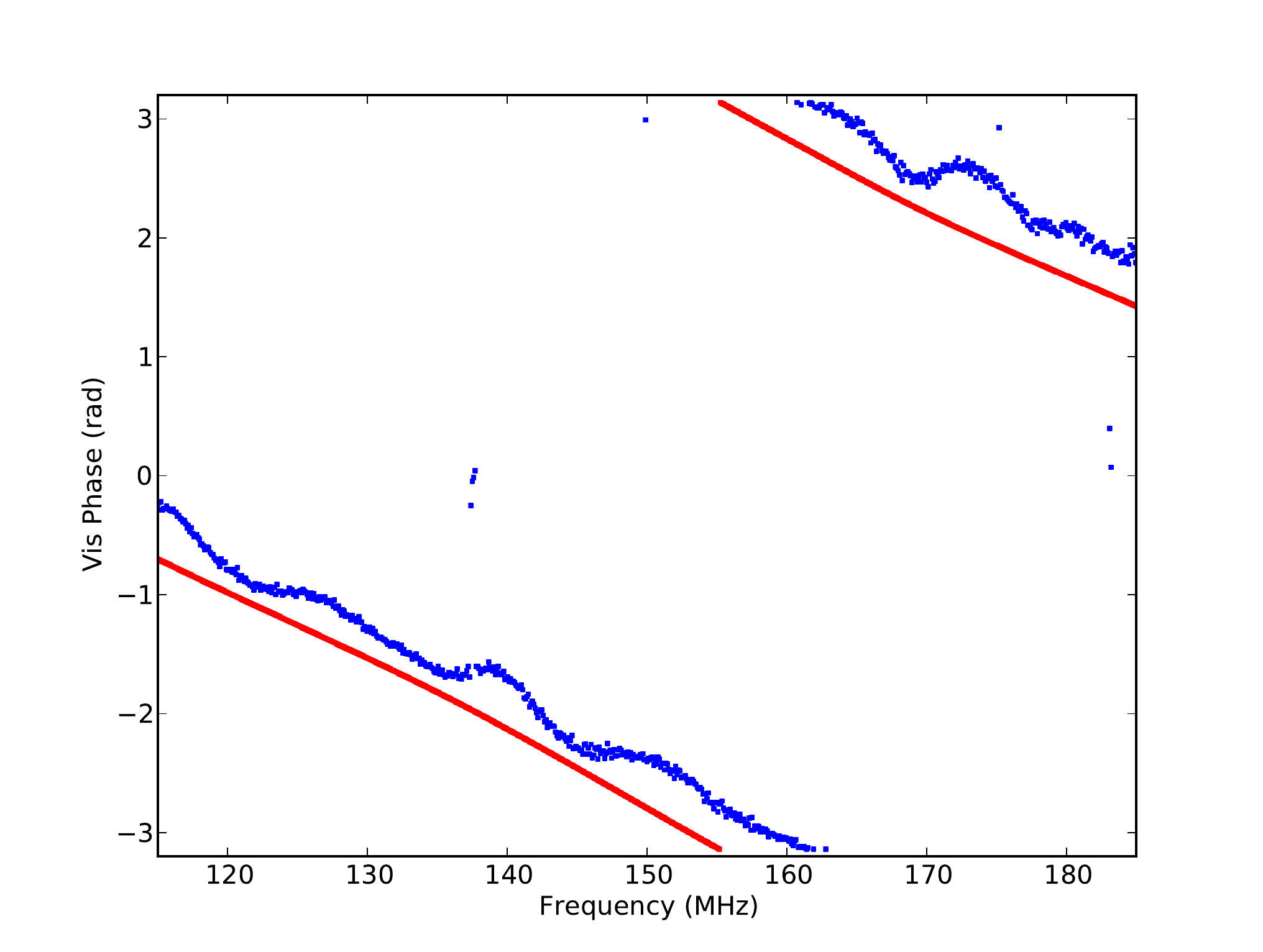}
\plottwo{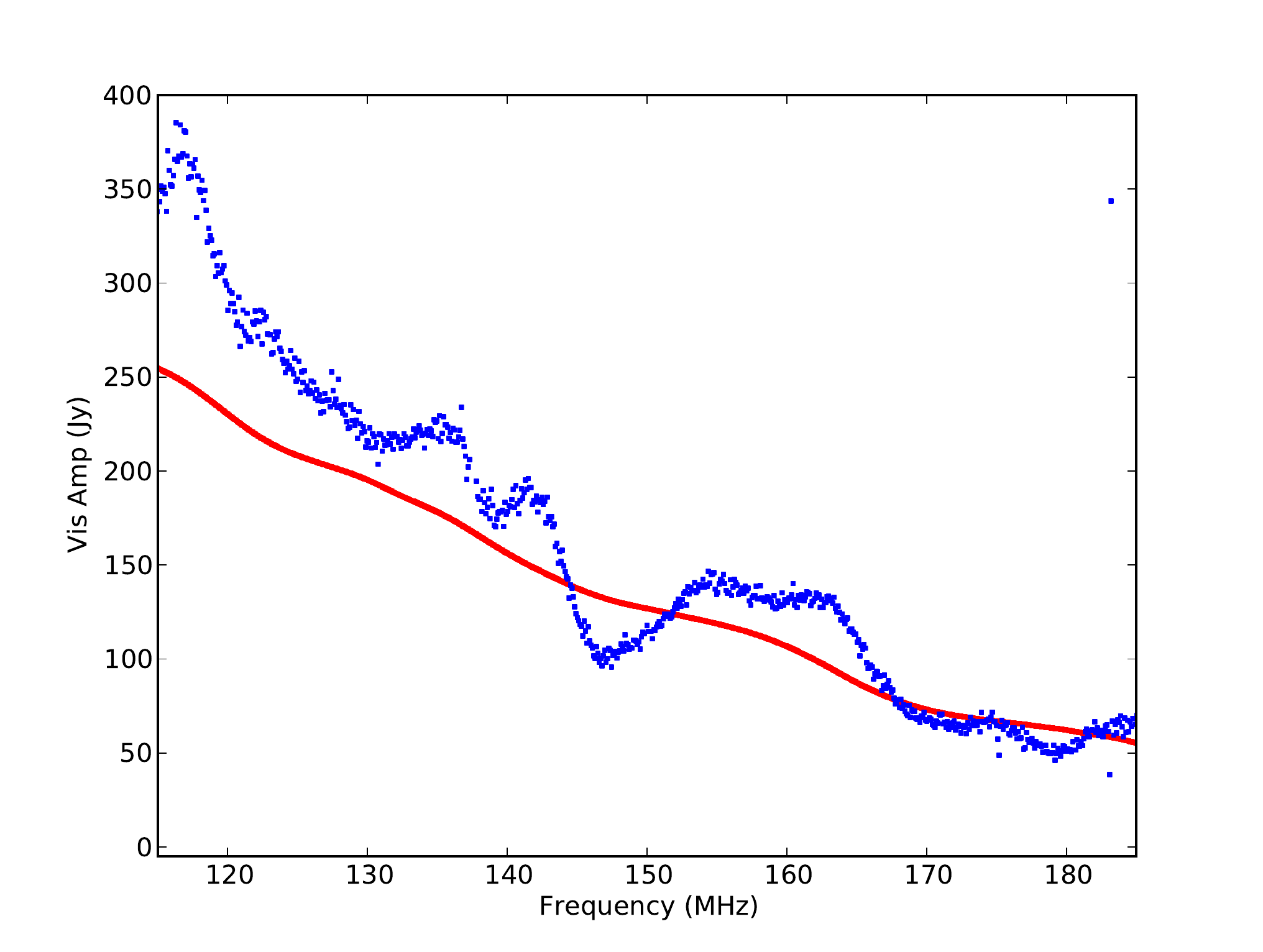}{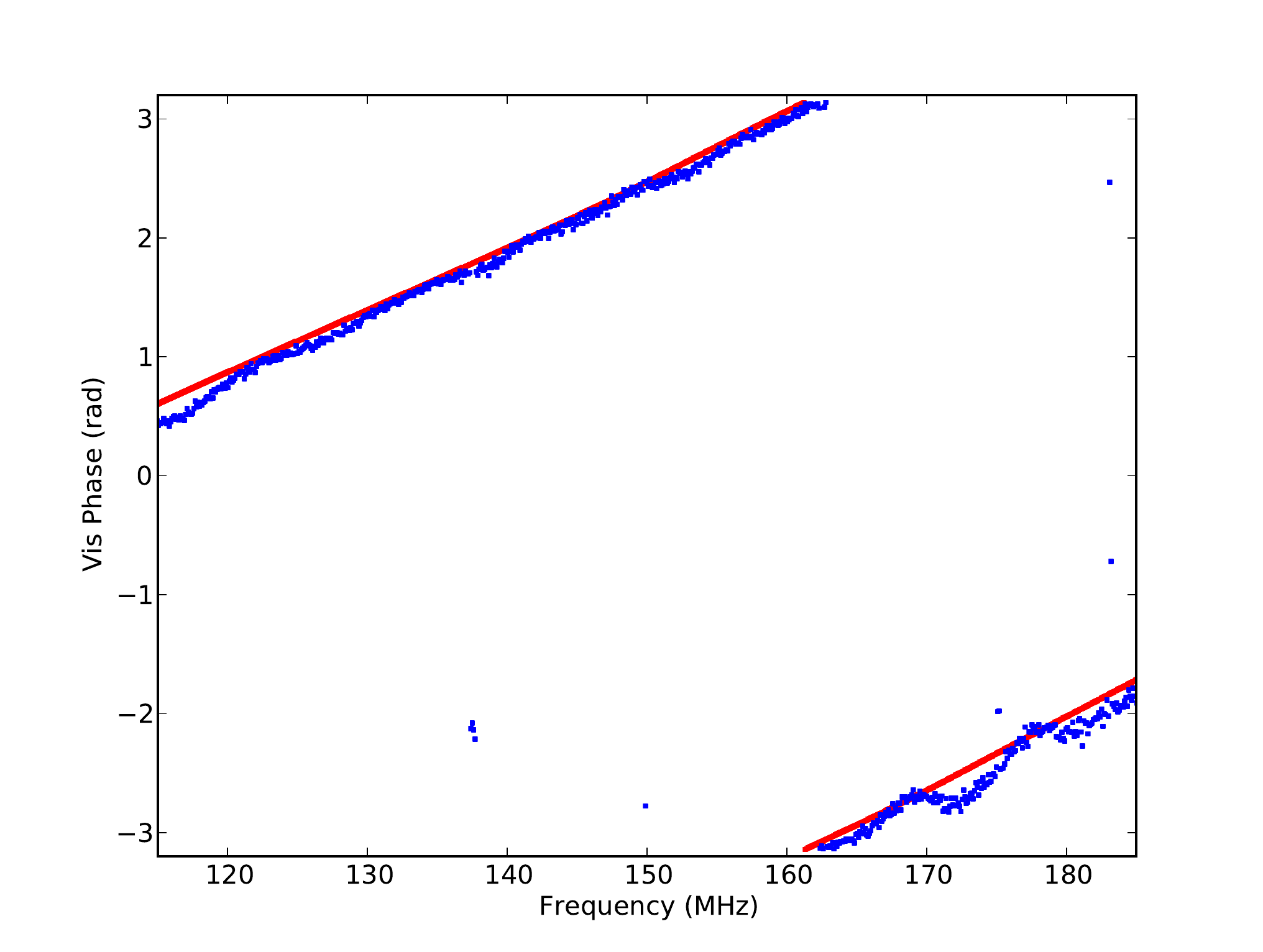}
\caption{Top (Left, Right): Amplitude and phase spectra on an east-west 29 m baseline for one record at transit of Fornax A. Blue shows the data. Red shows the GLEAM + Fornax A model. Middle: same, but for a 29 m baseline oriented at $-30^\circ$ with respect to north.  Bottom: same, but for a 29 m baseline oriented at $+30^\circ$ with respect to north.
}
\label{fig:f3}
\end{figure}

\begin{figure}
\plotone{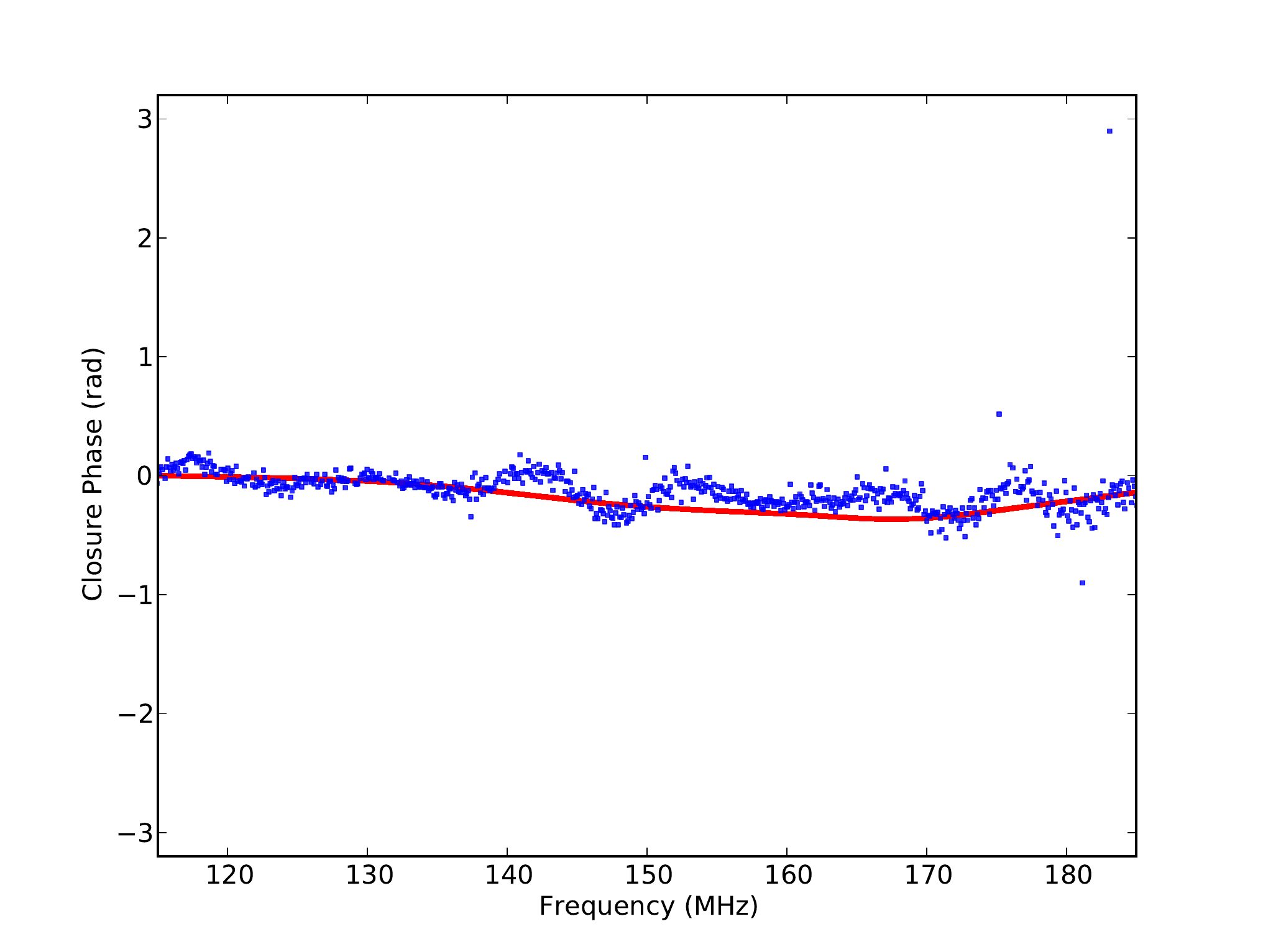}
\caption{Closure phase spectrum on a 29 m equilateral triad for the Fornax A field at transit. Blue shows the data, and red shows the GLEAM plus Fornax A model. 
}
\label{fig:f4}
\end{figure}

\begin{figure}
\includegraphics[scale=0.65]{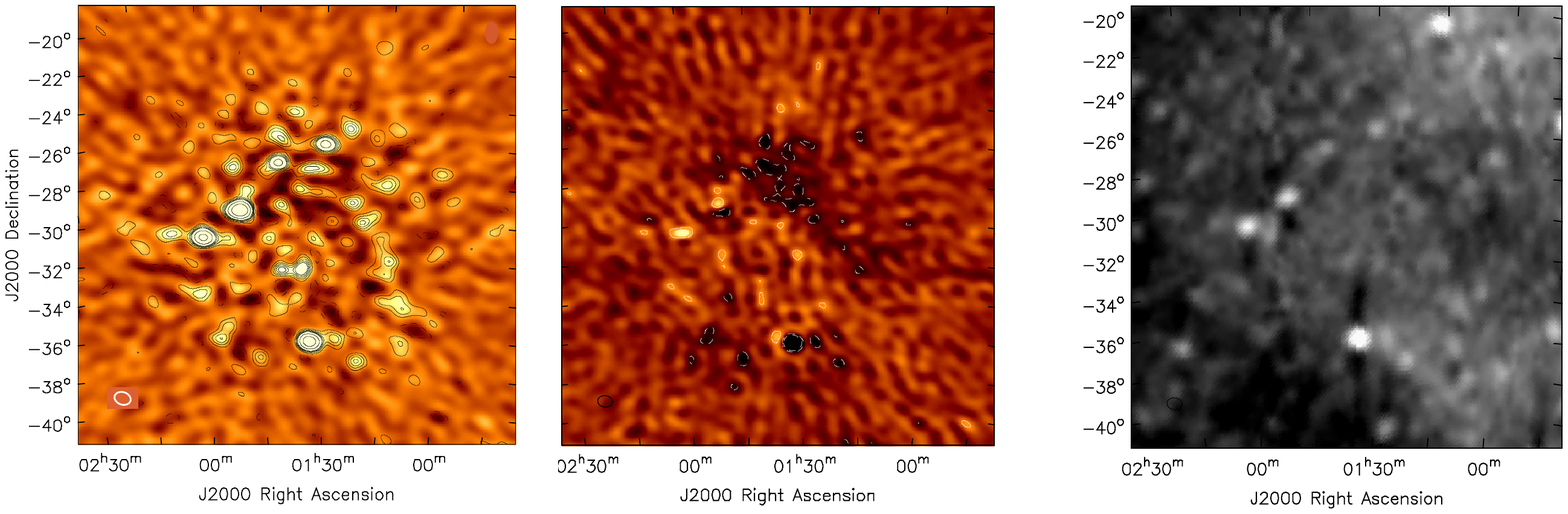}
\caption{Left: Image of the J0137-3042 field using 4~min of data over transit. The color scale shows the image made from the LST-binned IDR2 data from HERA, involving 18 days, using a multifrequency synthesis from 120 MHz to 180 MHz. The contours show the J0137 field GLEAM model, passed through the PRISim simulation and HERA telescope model, to generate visibilities, then imaged in CASA in the exact same way as for the real data. The contour levels are: -1.0, -0.5, 0.5, 1.0, 1.5, 2.0, 2.5, 5.0, 10.0 Jy beam$^{-1}$, and the resolution is $42' \times 33'$, PA = $65^\circ$. Dash contours are negative. The rms noise level outside the primary beam is 0.4~Jy~beam$^{-1}$. Middle: Difference image between model and data. The contour levels are the same as in the left image, to indicate quantitatively the relative magnitude of the residual features. Right: Image of the same field, but taken from the all-sky, total power image at 408 MHz of \citet{haslam, remazeilles15}.
}
\label{fig:f5}
\end{figure}

\begin{figure}
\plottwo{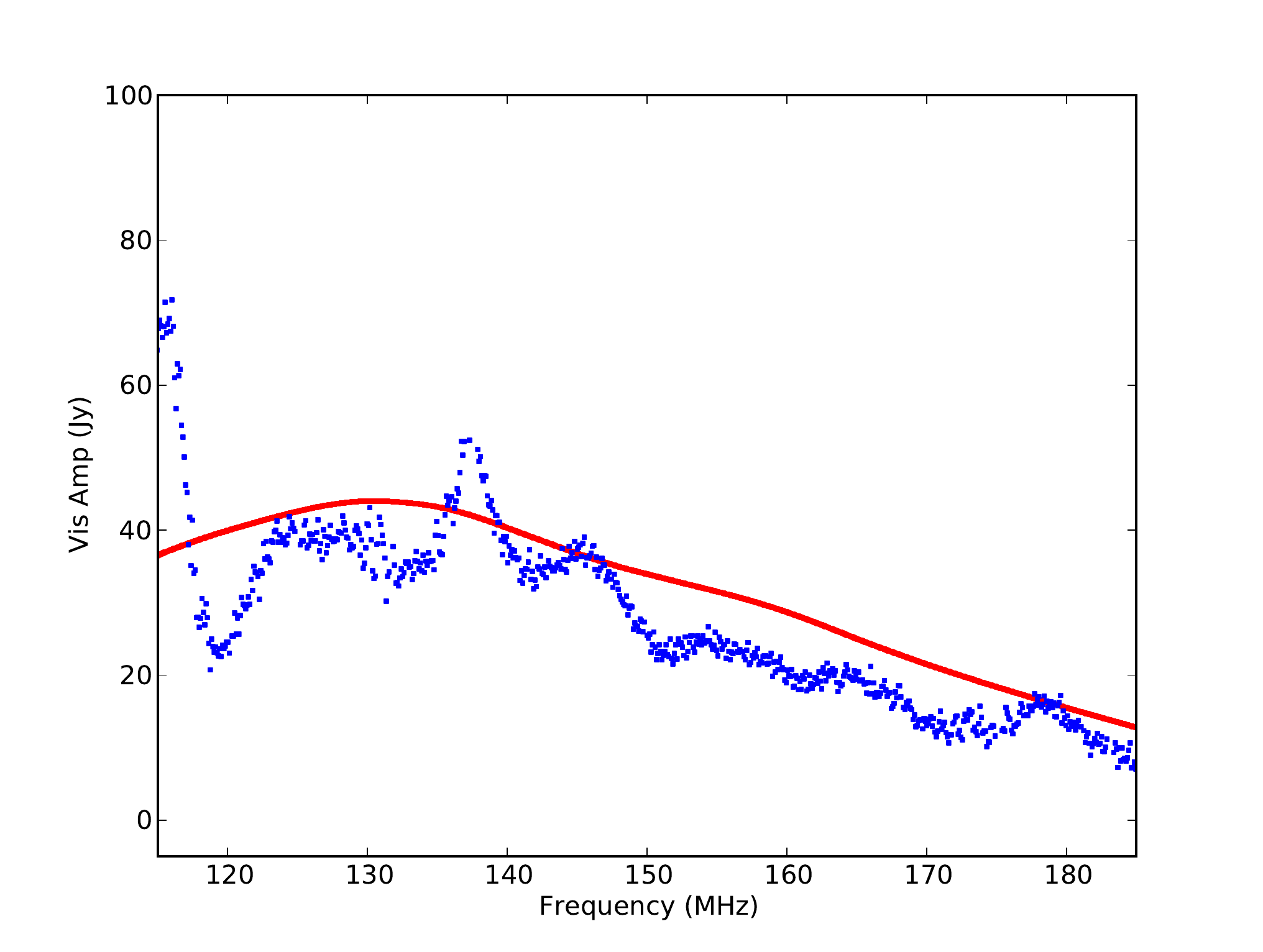}{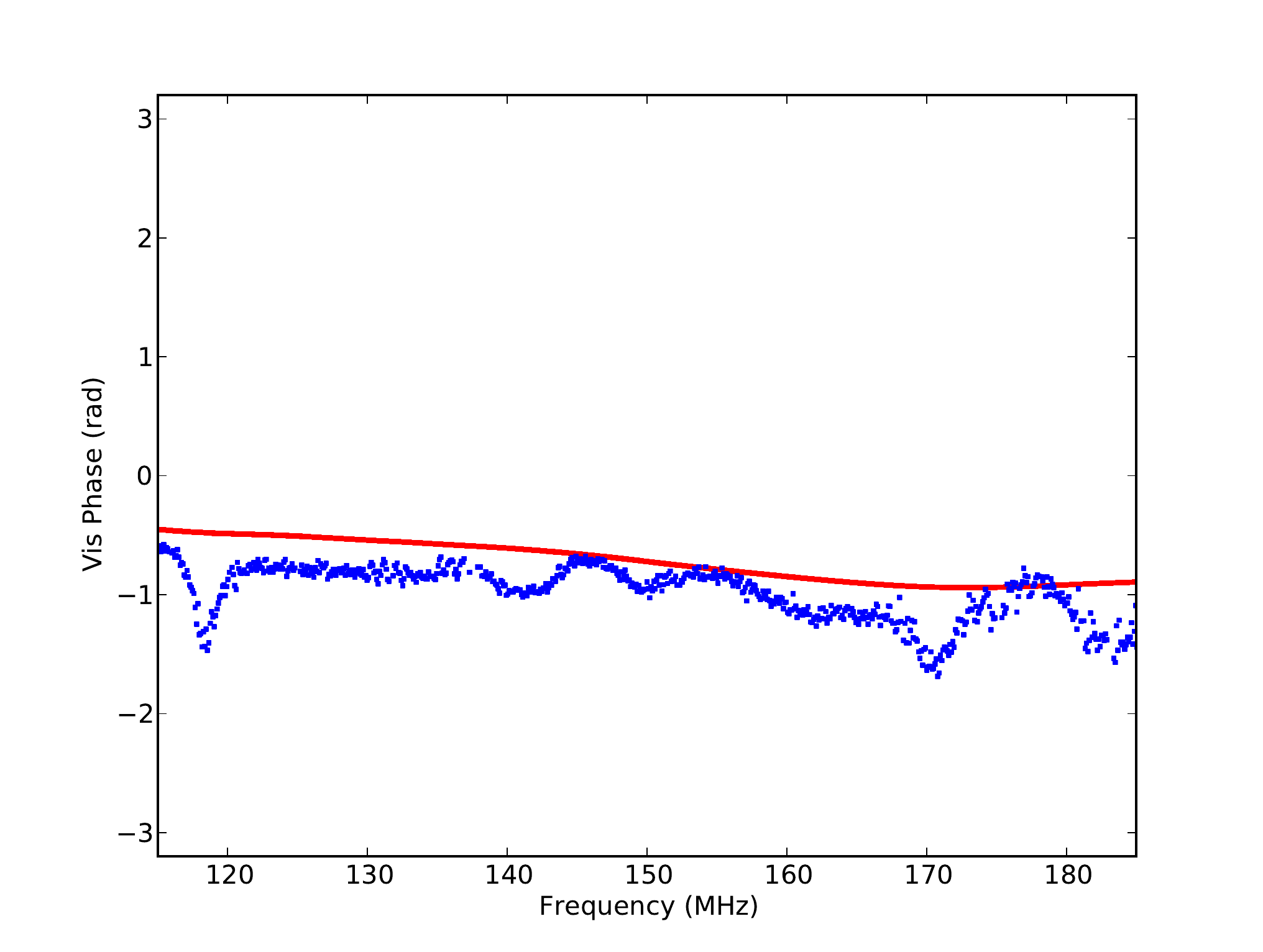}
\plottwo{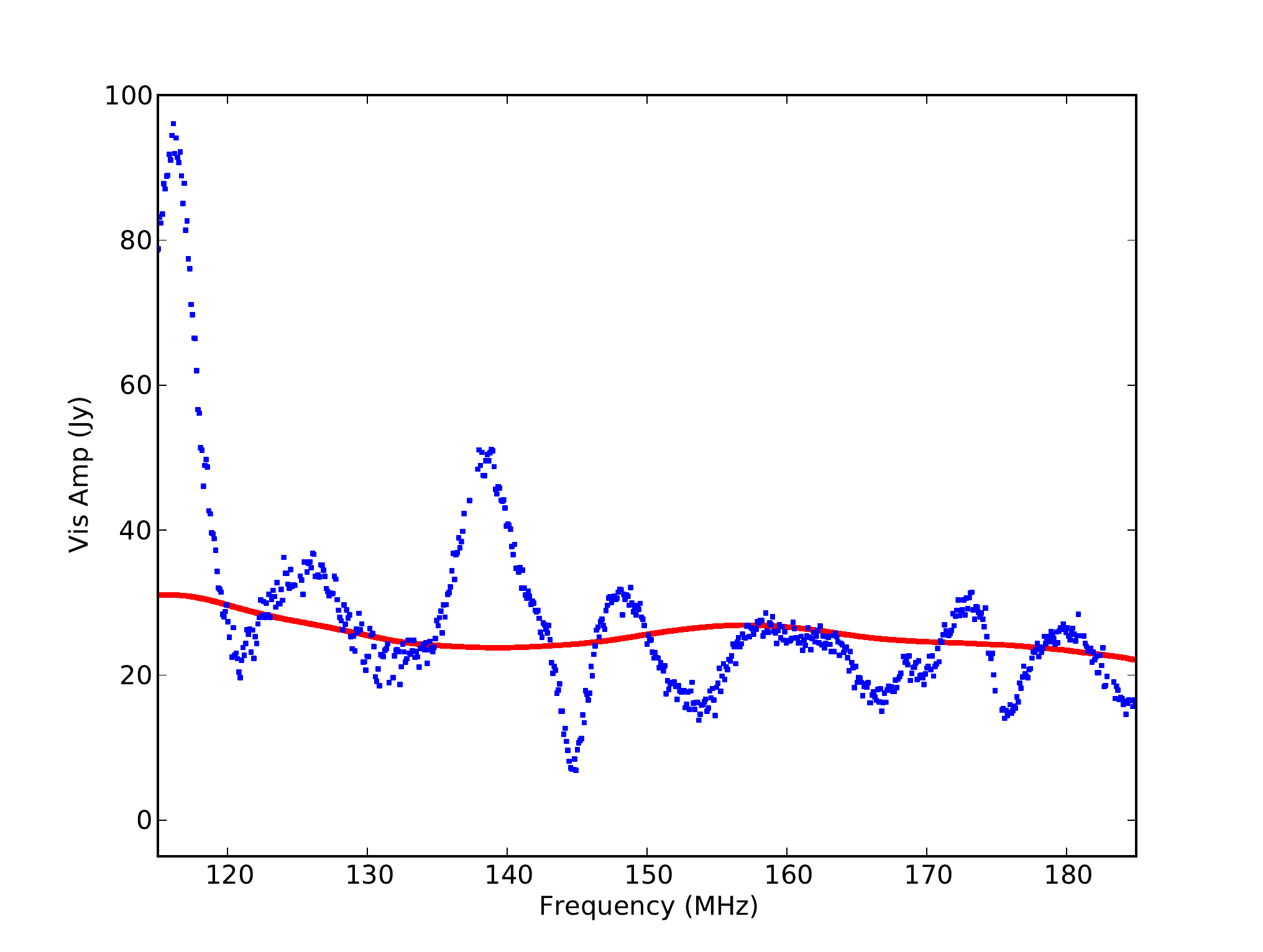}{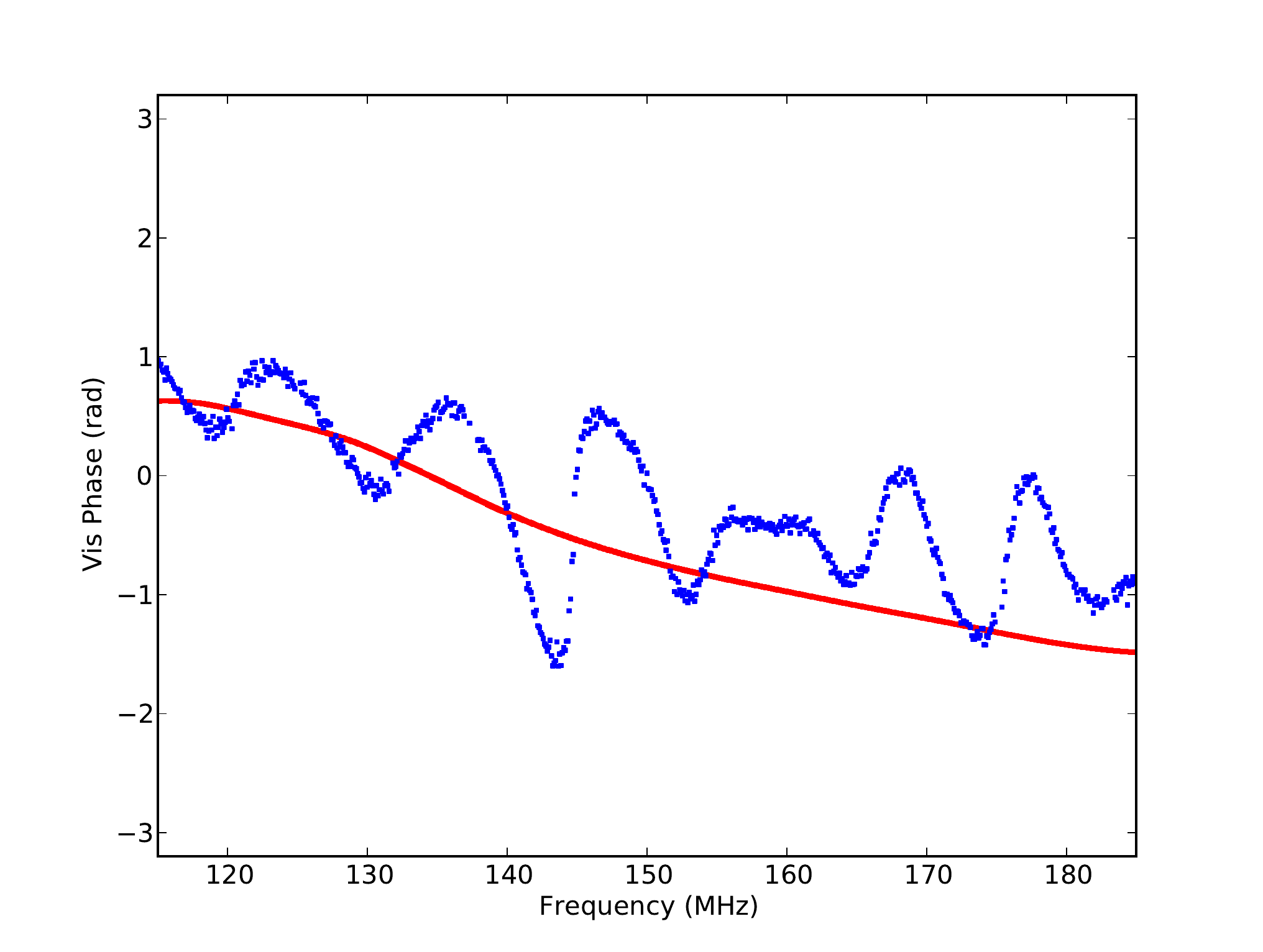}
\plottwo{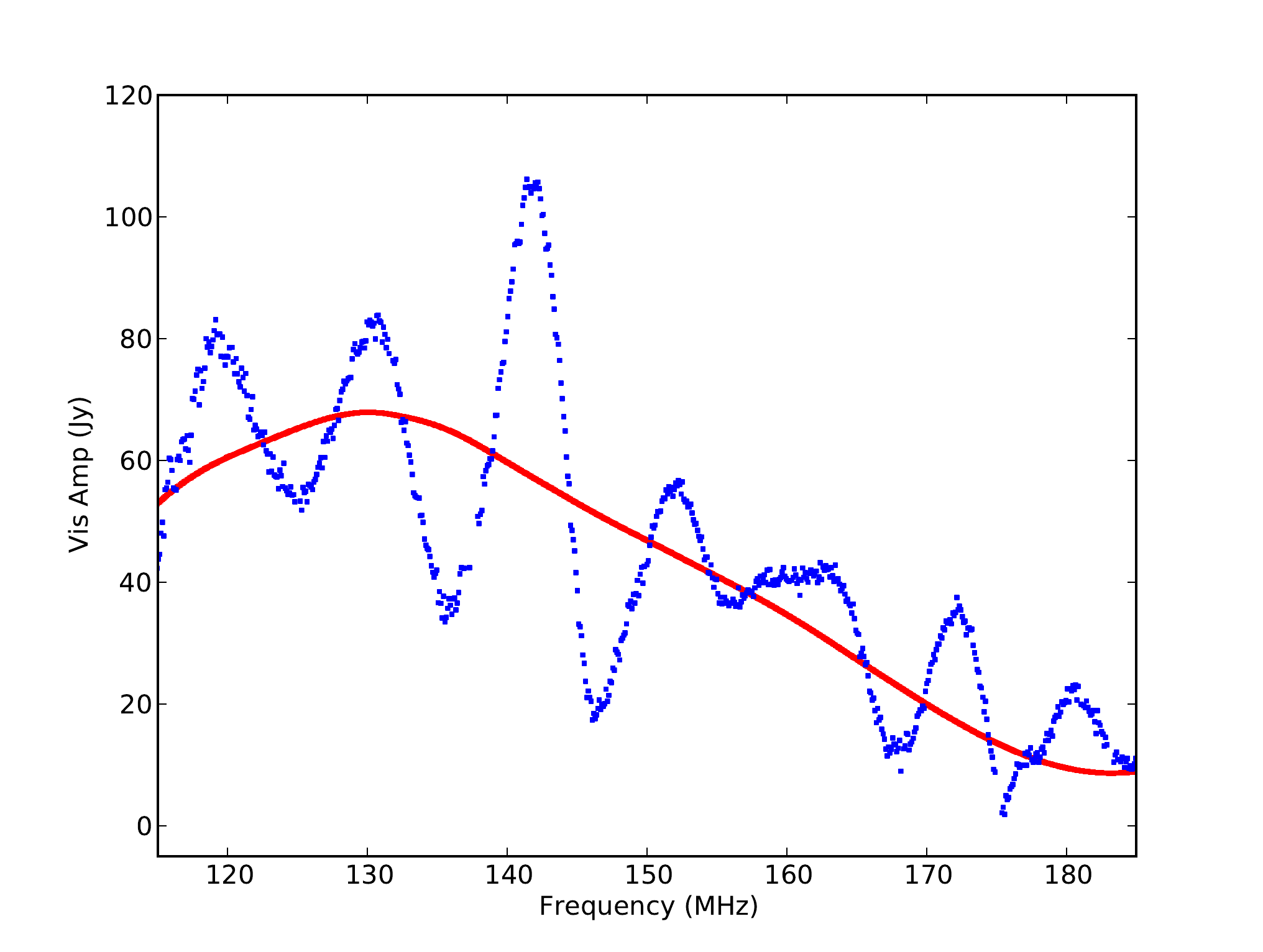}{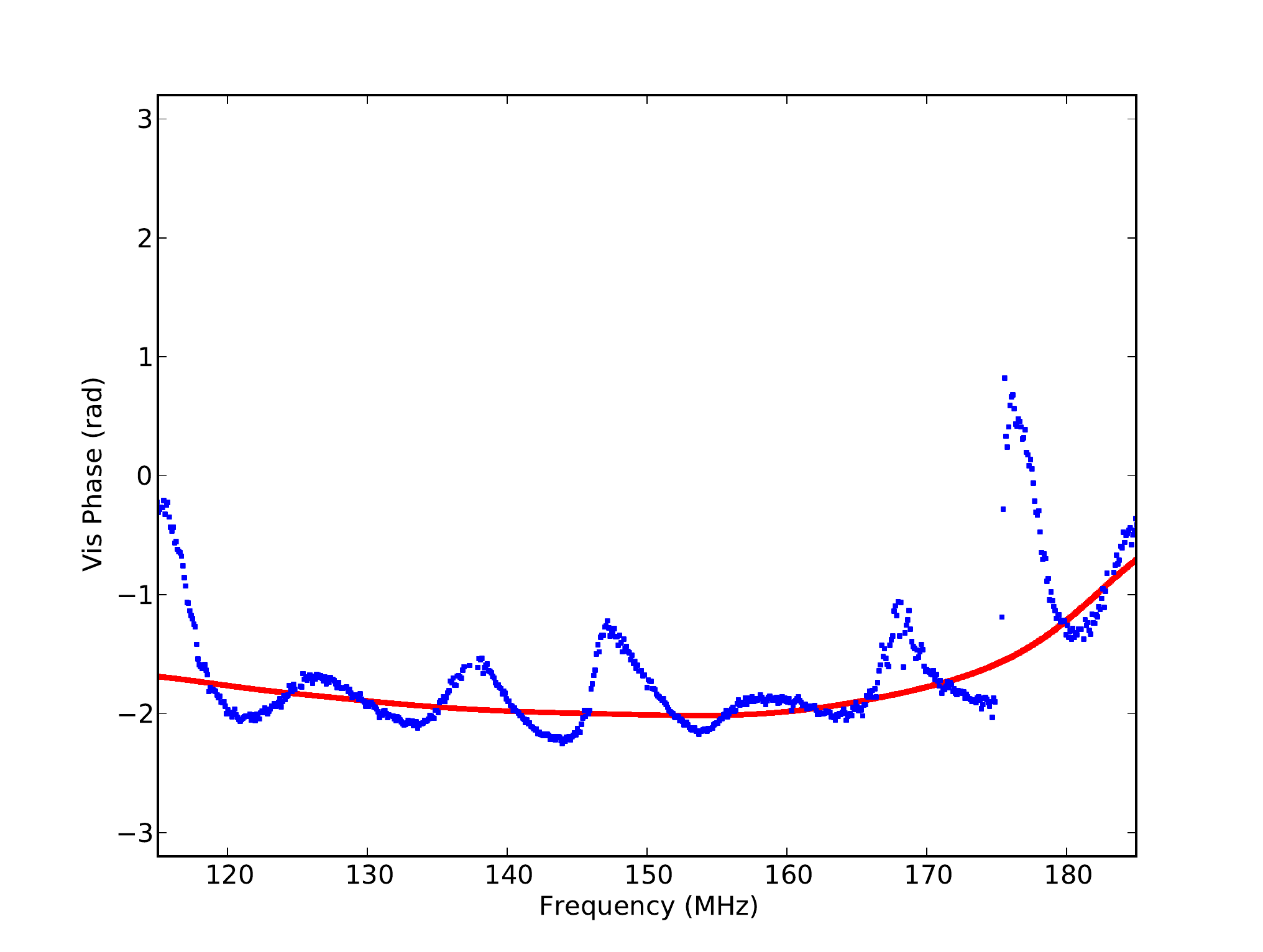}
\caption{Top (Left, Right): Amplitude and phase spectra on an east-west 29 m baseline for one record at transit of the J0137 field. Red shows a GLEAM point source only model. Blue shows the calibrated data. Middle: same, but for a 29 m baseline oriented at $-30^\circ$ with respect to north.  Bottom: same, but for a 29 m baseline oriented at $+30^\circ$ with respect to north. }
\label{fig:f6}
\end{figure}

\begin{figure}
\plotone{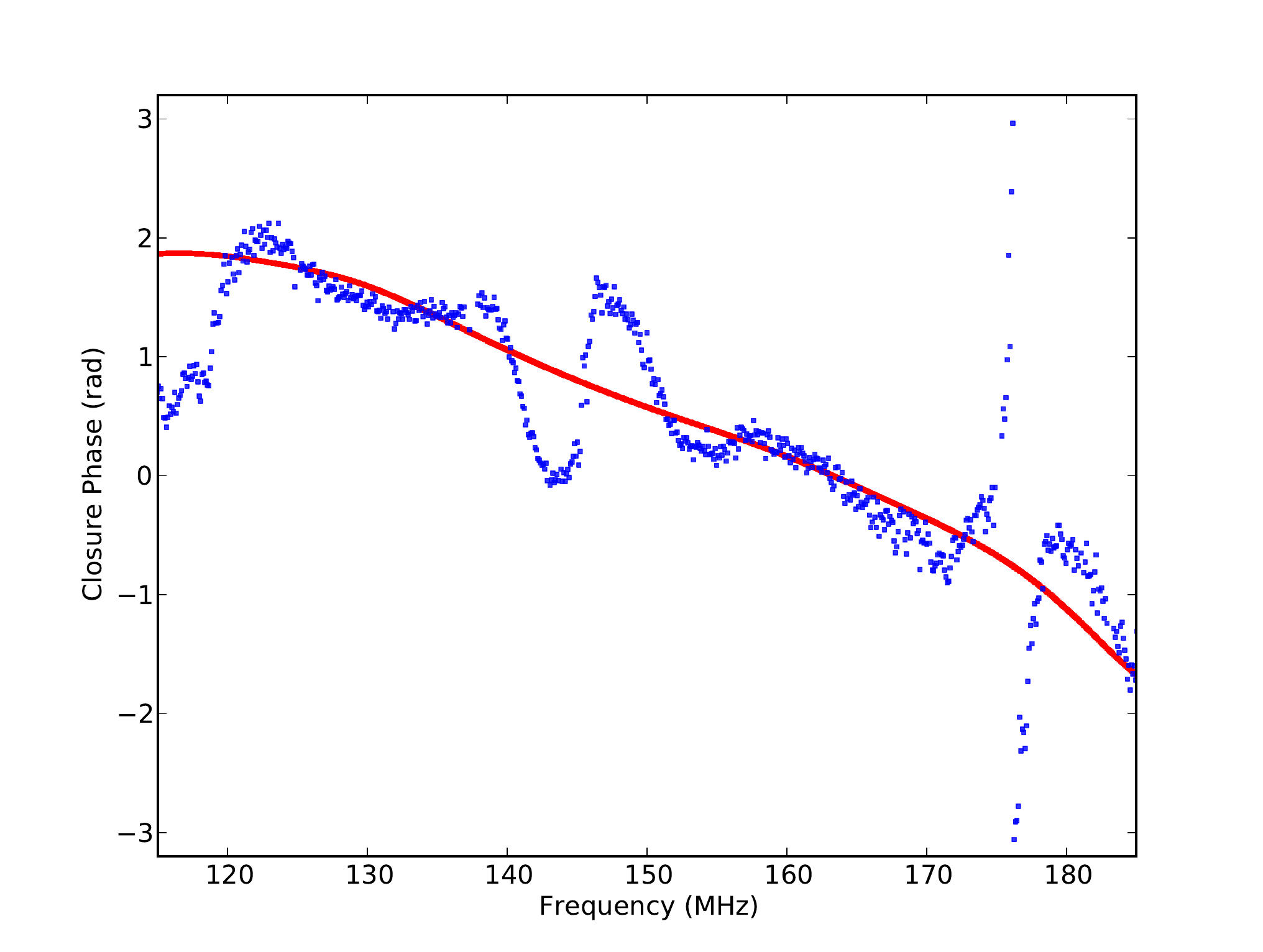}
\caption{Closure phase spectrum on a 29 m equilateral triad for the J0137 field at transit. Blue shows the data, and red shows the GLEAM point source model.
}
\label{fig:f7}
\end{figure}

The resulting images from the GLEAM source models demonstrate clearly the confusion limited imaging properties of a telescope such as HERA. However, the detail comparison of visibility spectra from GLEAM source-only models with the observed data shows a clear omission of structure in the models, corresponding to a $\sim 10$MHz scale wavy pattern in the data that is not seen in the model. In this section, we investigate this extra spectral structure, and conclude that it is likely due to diffuse Galactic emission filling the far sidelobes of the primary beam, and not captured in a GLEAM-only model.  Figure~\ref{fig:f1}, shows that parts of the bright Galactic plane are always above the horizon for HERA, even for the coldest regions of the sky at zenith.

\begin{figure}
\plotone{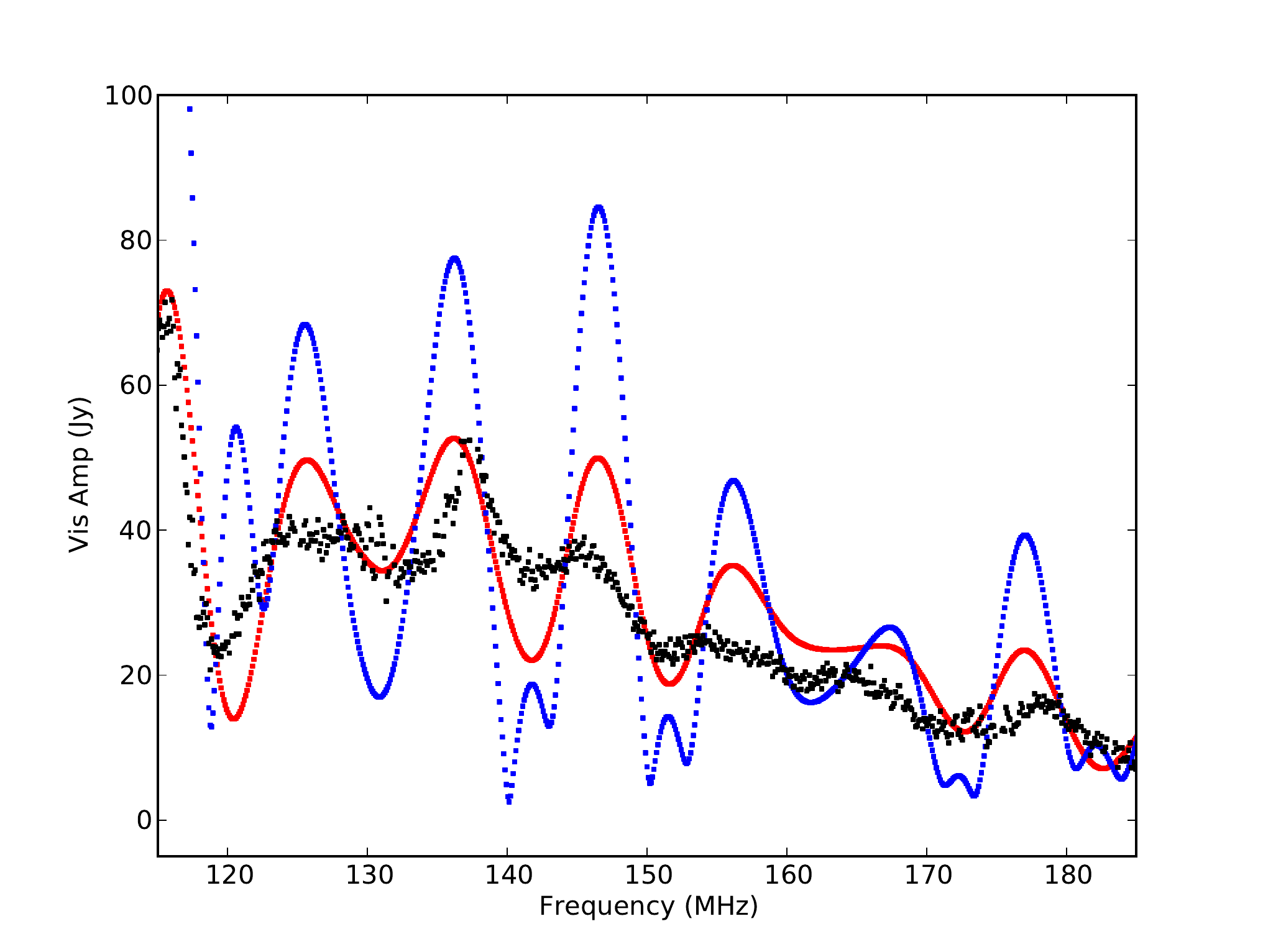}
\caption{The blue curve shows visibility spectra on a 29m east-west baseline for a model that includes the GLEAM point sources for the J0137 field, plus a diffuse all-sky model, both weighted by the wide field telescope primary beam model of \citet{fag17}. The red curve shows the same models, but with the diffuse all-sky models scaled down by an arbitrary factor three.
The black curve is the observed data. 
}
\label{fig:f8}
\end{figure}

We have generated a model that includes the GLEAM sources for the J0137-3042 field as per section 3.2, and add a diffuse all-sky model generated from the \citet{costa08}  analysis of low resolution, low frequency all-sky imaging. We then multiply the models by the full-sky power-response of the HERA antenna based on the electromagnetic modeling in \citet{fag17}. We present two models and the data in Figure~\ref{fig:f8}. One model includes the nominal diffuse sky model at full strength (blue), and the second scales the diffuse model down by a factor of three (red). The factor three down-scaling is arbitrary, simply to show the behaviour of the resulting spectra with changes in the strength of the diffuse emission model. 

We see that the full strength diffuse model over-predicts the observed spectral fluctuations by a large factor, while the fluctuations in the factor three down-scaled down diffuse model are closer to those observed. This result demostrates that, yes the diffuse model addition is likely the cause of the extra spectral structure, but that, unfortunately, our best current knowledge of the far-field beam structure and diffuse sky does not reproduce the observed visibilty structure with any accuracy.

Figure 9 shows the GLEAM sources plus the factor three down-scaled model compared to the measured visibility spectra in amplitude, phase, and closure phase, for a 29m east-west baseline. The fluctuations in phase and amplitude are of similar magnitude, with similar locations of maxima and minima with frequency, although there remain substantial differences in detail. 

Figure~\ref{fig:f10}a shows a comparison of the GLEAM plus the scaled diffuse sky model for a 29m and a 44m east-west baseline. The behaviour is as expected, in that the longer baseline has a higher frequency spectral structure, and is lower amplitude, as the diffuse emission becomes resolved. Extending such an analysis to much longer baselines becomes problematic, since the spectral structure due to the point sources themselves becomes the dominant effect in the measured visibilities.

Figure~\ref{fig:f10}b shows a comparison of the model spectra for three 29m baselines of different orientation. Substantial amplitude differences are seen between the different baseline orientation. Such differences are expected, as the visibility fringe projects along, or transverse to, large scale sky structure. Figure~\ref{fig:f1} shows that, for this particular field, the Galactic plane skirts the entire hemispheric rim, from east-to-north-to-west, with the fainter outer galaxy above the horizon to the east, and the brighter inner galaxy just below the horizon to the west, with parts of the thicker disk in the inner galaxy extending above the horizon. Hence, it is not easy to predict which baseline orientation will have the largest amplitude. 

Overall, the evidence suggests that the extra spectral structure in visibility spectra not captured in the GLEAM-only model, but seen in the data, is due to diffuse Galactic emission, dominated by the bright Galactic plane in the far sidelobes of the primary beam. The challenges of generating a full sky model including the diffuse Galactic emission plus the extragalactic and Galactic point sources are manifold. First is knowledge of the diffuse emission, and its broad band spectral distribution. Second is accurate knowledge of the primary beam as a function of frequency at large zenith angles. And third is the inevitable double-counting of the plethora of faint extragalactic and Galactic point sources that appear as a diffuse component at low resolution. This latter effect is particularly problematic in the modeling effort.

Accurate measurement of the far-field primary beam pattern is a severe challenge for non-pointing (zenith) instruments, such as HERA, although techniques using celestial sources are being explored \citep{nun19, pober12}. Having an antenna that can point and track over the sky would be clearly advantageous to perform wide-field holography and hence provide a much better measurement of the wide field primary beam (see \citet{napier99}). We discuss some of the implications of this spectral structure, and mitigation techniques, below. 

\section{Discussion}

We have made a detailed comparison of HERA-50 visibility spectra and images with sky models generated from the GLEAM survey, plus a HERA telescope model, processed through the PRISim simulator. These models provide an important comparison to the data in our closure phase power spectral analysis, in search of the HI 21cm signal from cosmic reionization \citep{nithya19a}. 

We emphasize that the modeling used in the closure phase power spectral analysis is not required for calibration nor source subtraction, as is required in other power spectral techniques that rely on antenna-based array calibration.  The closure phases are independent of simple antenna based calibration terms, ie. single antenna-based complex gains per frequency per antenna. Further, the delay spectrum approach limits the smooth spectrum foregrounds to low delay modes, and hence is amenable to delay filtering in the power spectrum  in order to isolate the HI signal at larger delays \citep{parsons14, nithya19a}.

For the closure phase power spectral analysis, we require sky models simply to check the scaling of the relative magnitude of effects, such as continuum emission, noise, and the HI 21cm signal, for comparison to the measured power spectra derived from closure phase spectra. Hence, the required accuracy of the models is much relaxed relative to other techniques. For instance, given the statistical significance of the eventual HI detection with HERA will be at most $\sim 5$ to 10$\sigma$ for HERA, the modeling accuracy needs only be good to roughly the 10\% level,  as a guide for interpreting the closure phase power spectral results \citep{thya18, nithya19a}. 

The results indicate that the images derived from uv-data generated using PRISim and a GLEAM survey sky model, plus the HERA-50 telescope model, match the images derived from the real data down to the confusion limit of the telescope of $\sim 0.4$ Jy beam$^{-1}$. The measured noise level in the field is orders of magnitude above the theoretical thermal noise, even for short integrations. HERA-50 is deeply in-beam source confusion limited due to the low spatial resolution and high source areal density at low frequency. Analysis of a residual image between sky and model images, with the total power image of the field, suggests diffuse sky structure at around the confusion level, that is not represented in a point source only model.

We conclude that, for the resulting broad-band images, the GLEAM sky model, plus the PRISim implementation of the telescope model, is a good representation of the data at the strongly confusion-limited level that can be measured with the HERA array.  Correspondingly, the calibrated HERA-50 data generate an image that matches, to the confusion level, what is predicted for the sky surface brightness distribution in these high latitude fields.

For the visibility spectra in amplitude, phase, and closure phase, there is good agreement of the broad structure across the spectral range, but the measurements themselves show a roughly sinusoidal pattern on scales of $\sim 10$ MHz which is not reproduced in the GLEAM-only model. Adding an all-sky, diffuse emission component, dominated by the Galactic plane at large zenith angles, produces a plausible explanation for this spectral structure. Detailed modeling of this very wide field sky emission remains problematic due to uncertainties in both the primary beam model and diffuse Galactic emission models, both as a function of frequency, as well as `double counting' of the fainter extragalactic sources that fill the sky.

How will the unmodeled extra spectral structure affect the HERA search for the HI 21cm signal from cosmic reionization? This question has been considered in \citet{kern19} and \citet{nithya19a}, which we briefly summarize.

First is the effect on sky calibration using external models. Any unmodeled structure that is in the data will propagate through the bandpass calibration and lead to errors. This effect has been considered by a number of authors, including \citet{byrne19, li19, barry16, vanweeren16, ewall17}. Most recently, \citep{kern19} show that the unmodeled spectral structure leads to a peak in the amplitudes in the delay transform at around $\sim 200$ nanosec, implying potential contamination of the measured HI 21cm power spectrum using the delay spectrum approach at low wavenumber $k \sim 0.1$ Mpc$^{-1}$. Fortunately, delay spectrum searches for HI 21cm emission have thus far relied on analysis at wavenumbers, $k \ge 0.2$ Mpc$^{-1}$ \citep{parsons14}.

Redundant baseline calibration may be immune to this phenomenon, since it relies on the measured visibilities themselves, but ultimately, remaining degeneracies require that an average system bandpass be derived based on sky models \citep{dillon16, zheng17, wieringa92, rammarthi14}, and hence the problem is not absent \citep{kern19, orosz19}.

Self-calibration using models derived from the data itself may perform somewhat better than using {\sl a priori} sky and telescope models. However, restoring very broad diffuse emission in an interferometric image is a challenge \citep{rau11}, and impossible in cases where the structure is much larger than the shortest spacing of the array. One might consider a process of only using long baselines to calibrate all baselines, since these are less affected by diffuse emission. However, this technique has its own drawbacks, such as higher frequency residuals across a visibility spectrum due to calibration errors, given the longer baselines involved \citep{ewall17, patil16}. 

In terms of the closure phase delay spectrum approach, calibration is not an issue, since the technique employs uncalibrated data. However, any spectral structure on these scales will also show up at similarly low wavenumbers in the power spectrum. If this structure is not paralleled in the modeling, then a comparison of the measured power in the data vs. the model at low wave numbers will not be appropriate \citep{nithya19a}.

\acknowledgments
The National Radio Astronomy Observatory is a facility of the National Science Foundation operated under cooperative agreement by Associated Universities, Inc.. This  material  is  based  upon  work  supported  by  the  National Science  Foundation  under  Grant  Nos.  1636646  and 1836019 and institutional support from the HERA collaboration partners. This research is funded in part by the Gordon and Betty Moore Foundation. HERA is hosted by the South African Radio Astronomy Observatory, which is a facility of the National Research Foundation, an agency of the Department of Science and Technology. We gratefully acknowledge the support of NVIDIA Corporation with the donation of the Titan X GPU used for this research. GB acknowledges funding from the INAF PRIN-SKA 2017 project 1.05.01.88.04 (FORECaST), support from the Ministero degli Affari Esteri della Cooperazione Internazionale - Direzione Generale per la Promozione del Sistema Paese Progetto di Grande Rilevanza ZA18GR02 and the National Research Foundation of South Africa (Grant Number 113121) as part of the ISARP RADIOSKY2020 Joint Research Scheme. CC, GB and SM acknowledge support from the Royal Society and the Newton Fund under grant NA150184. This work is based on research supported in part by the National Research Foundation of South Africa (grant No. 103424).

\clearpage
\newpage

\begin{figure}
\includegraphics[scale=0.28]{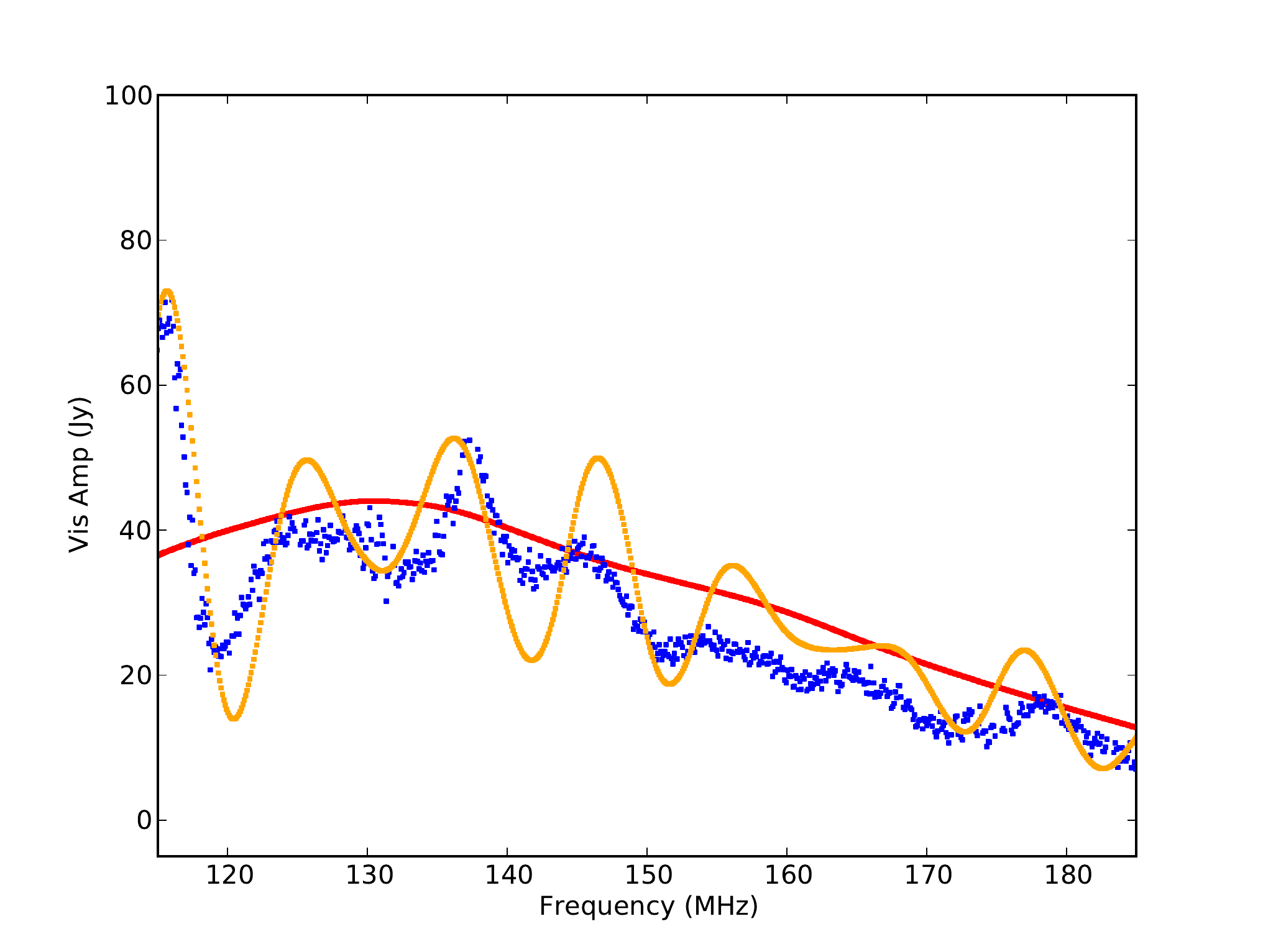}
\includegraphics[scale=0.28]{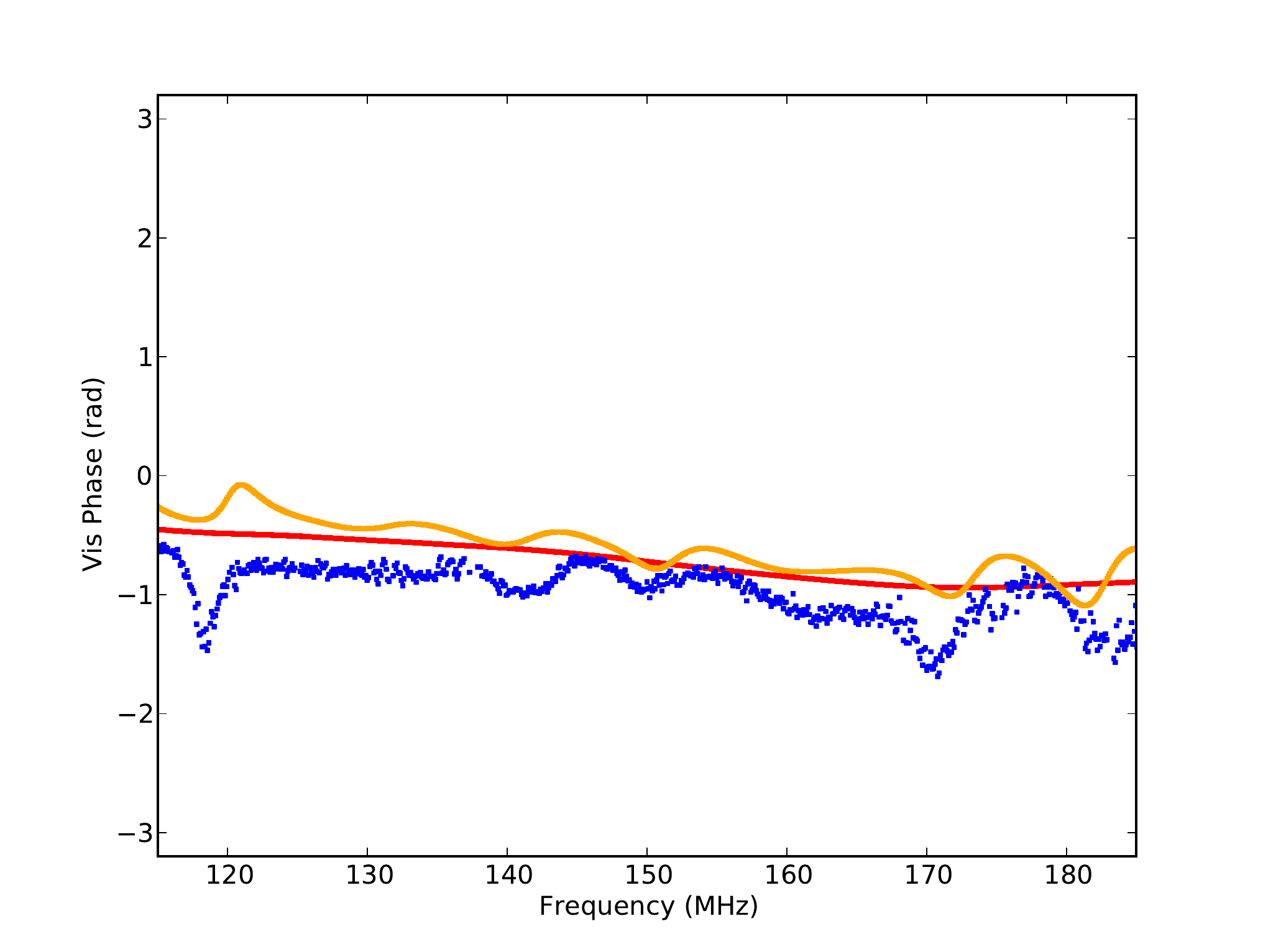}
\includegraphics[scale=0.28]{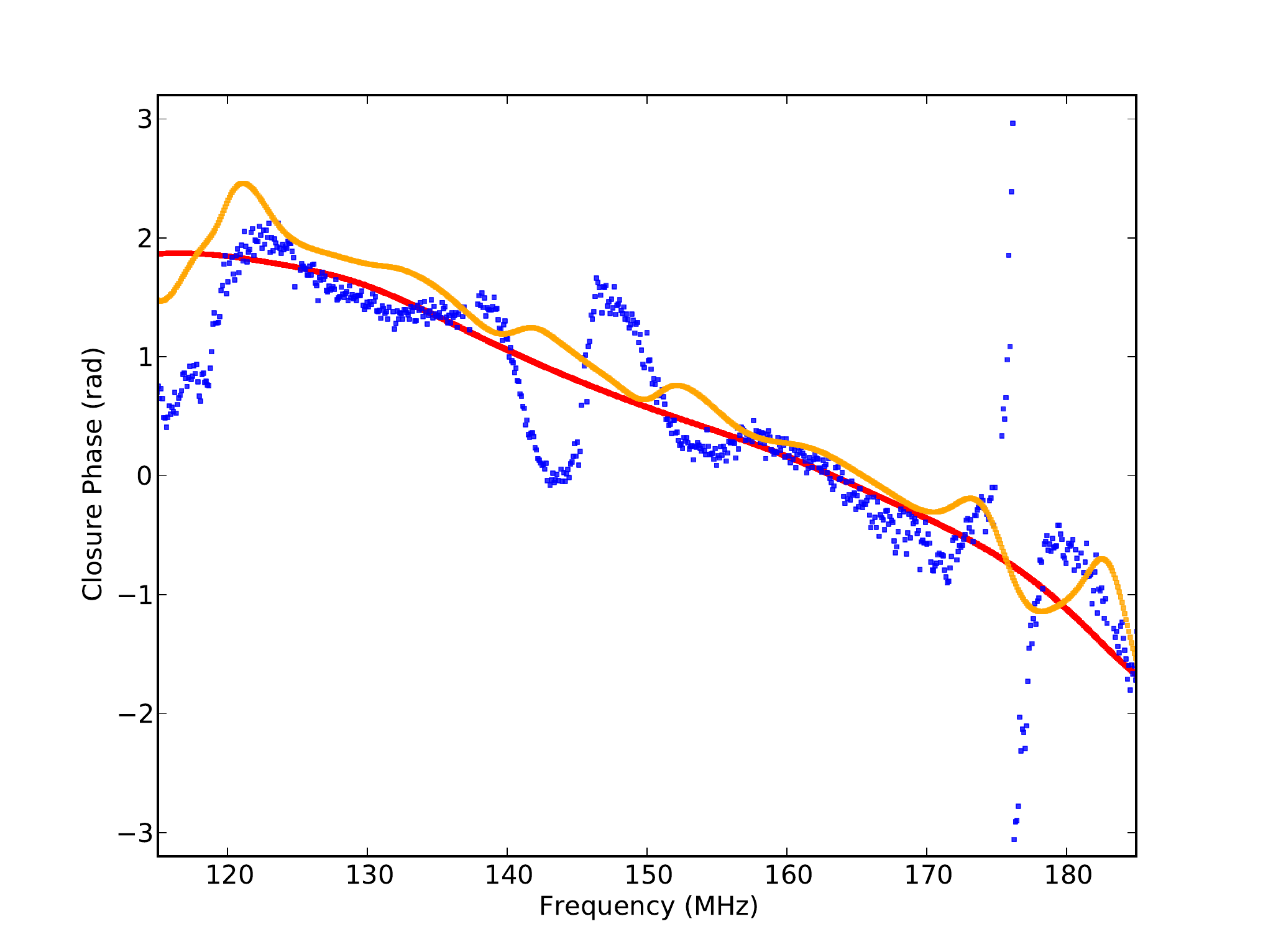}
\caption{Left: Visibility amplitude spectrum for a 29m east-west baseline in the J0137-3042 field. Blue is the HERA data, and red shows the model including only the GLEAM sources. The orange shows a model including the GLEAM sources, plus a factor three scaled-down all-sky diffuse model, as discussed in section 4. Center: same, but showing phase. Right: Same, but for the closure phase spectrum. 
}
\label{fig:f9}
\end{figure}

\begin{figure}
\plottwo{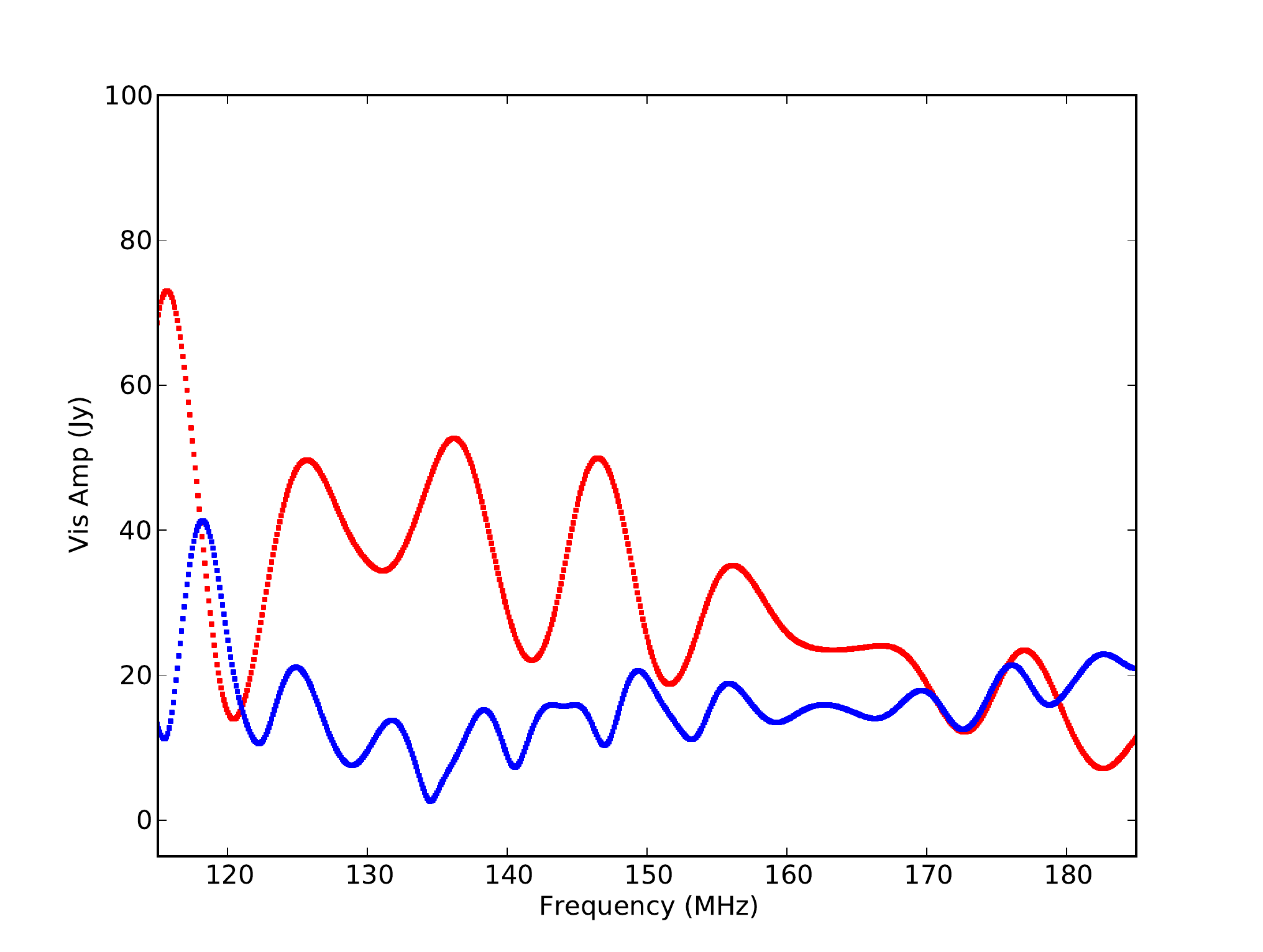}{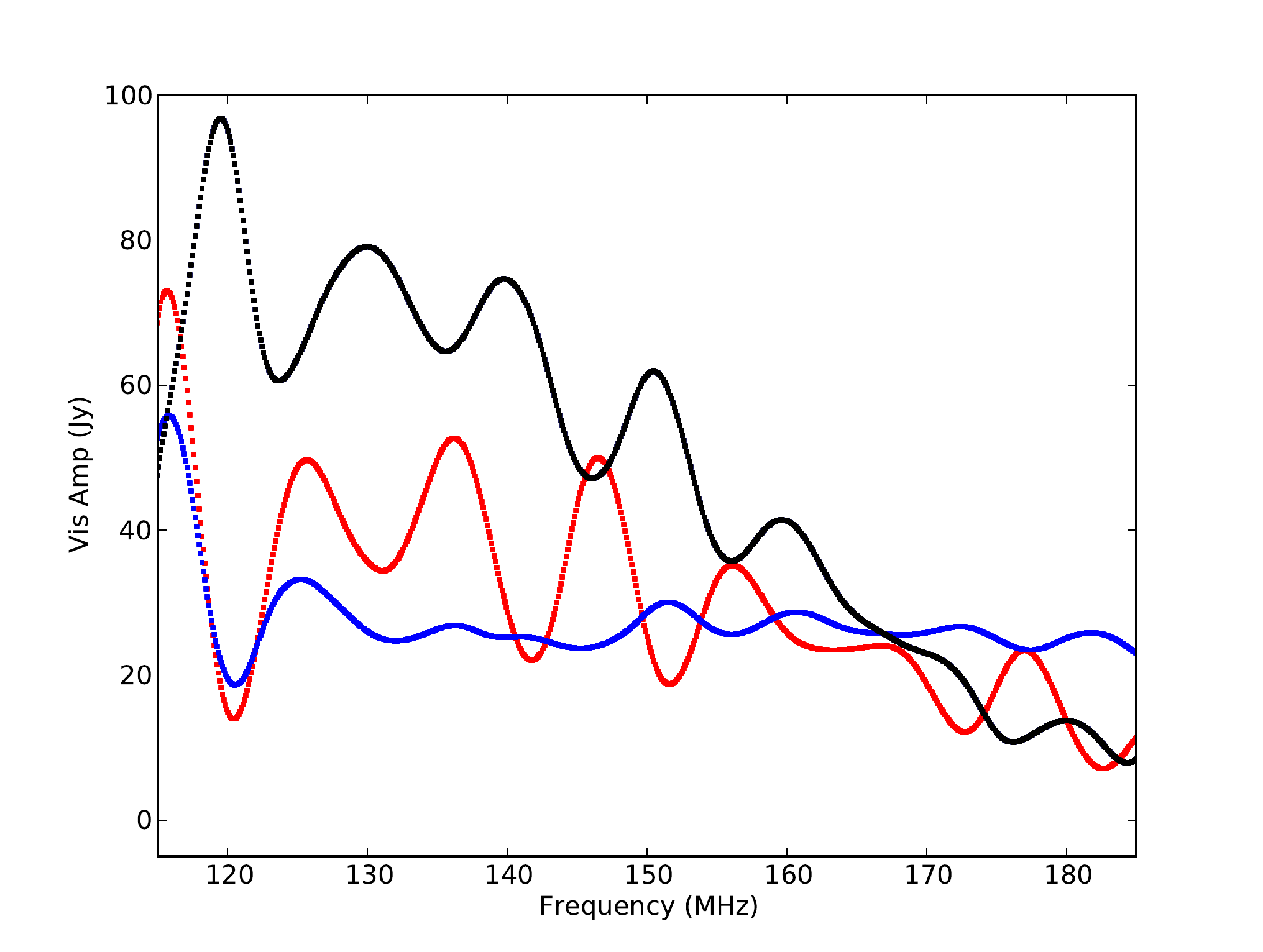}
\caption{Left: Red is the visibility amplitude spectrum for the GLEAM plus the scaled-down diffuse sky model (as show in figure 8) for a 29m baseline. Blue is the same, but for a 44m baseline. Right: Red shows the visibility amplitude spectrum for  the GLEAM plus the scaled diffuse sky model (as show in figure 8) for an east-west 29m baseline. Black shows a 29m baseline oriented $30^\circ$ counterclockwise from North. Blue shows a 29m baseline at
$30^\circ$ clockwise from North
}
\label{fig:f10}
\end{figure}

\end{document}